\documentclass[prd, showpacs,preprintnumbers,amsmath,amssymb,nofootinbib,aps]{revtex4}
\usepackage{bm}
\usepackage{amsfonts}
\usepackage{mathrsfs}
\usepackage{graphicx}
\usepackage{amsmath}
\usepackage{float}
\usepackage{braket}

\begin{document}

\newcommand{\hhat}[1]{\hat {\hat{#1}}}
\newcommand{\pslash}[1]{#1\llap{\sl/}}
\newcommand{\kslash}[1]{\rlap{\sl/}#1}
\newcommand{\lab}[1]{}
\newcommand{\iref}[2]{}
\newcommand{\sto}[1]{\begin{center} \textit{#1} \end{center}}
\newcommand{\rf}[1]{{\color{blue}[\textit{#1}]}}
\newcommand{\eml}[1]{#1}
\newcommand{\el}[1]{\label{#1}}
\newcommand{\er}[1]{Eq.\eqref{#1}}
\newcommand{\df}[1]{\textbf{#1}}
\newcommand{\mdf}[1]{\pmb{#1}}
\newcommand{\ft}[1]{\footnote{#1}}
\newcommand{\n}[1]{$#1$}
\newcommand{\fals}[1]{$^\times$ #1}
\newcommand{\new}{{\color{red}$^{NEW}$ }}
\newcommand{\ci}[1]{}
\newcommand{\de}[1]{{\color{green}\underline{#1}}}
\newcommand{\ke}{\rangle}
\newcommand{\br}{\langle}
\newcommand{\lb}{\left(}
\newcommand{\rb}{\right)}
\newcommand{\lbk}{\left[}
\newcommand{\rbk}{\right]}
\newcommand{\blb}{\Big(}
\newcommand{\brb}{\Big)}
\newcommand{\nn}{\nonumber \\}
\newcommand{\p}{\partial}
\newcommand{\pd}[1]{\frac {\partial} {\partial #1}}
\newcommand{\cd}{\nabla}
\newcommand{\cc}{$>$}
\newcommand{\bqa}{\begin{eqnarray}}
\newcommand{\eqa}{\end{eqnarray}}
\newcommand{\bqe}{\begin{equation}}
\newcommand{\eqe}{\end{equation}}
\newcommand{\bay}[1]{\left(\begin{array}{#1}}
\newcommand{\eay}{\end{array}\right)}
\newcommand{\eg}{\textit{e.g.} }
\newcommand{\ie}{\textit{i.e.}, }
\newcommand{\iv}[1]{{#1}^{-1}}
\newcommand{\st}[1]{|#1\ke}
\newcommand{\at}[1]{{\Big|}_{#1}}
\newcommand{\zt}[1]{\texttt{#1}}
\newcommand{\non}{\nonumber}
\newcommand{\m}{\mu}
\def\xa{{\alpha}}
\def\xA{{\Alpha}}
\def\xb{{\beta}}
\def\xB{{\Beta}}
\def\xd{{\delta}}
\def\xD{{\Delta}}
\def\xe{{\epsilon}}
\def\xE{{\Epsilon}}
\def\xve{{\varepsilon}}
\def\xg{{\gamma}}
\def\xG{{\Gamma}}
\def\xk{{\kappa}}
\def\xK{{\Kappa}}
\def\xl{{\lambda}}
\def\xL{{\Lambda}}
\def\xo{{\omega}}
\def\xO{{\Omega}}
\def\xvp{{\varphi}}
\def\xs{{\sigma}}
\def\xS{{\Sigma}}
\def\xt{{\theta}}
\def\xvt{{\vartheta}}
\def\xT{{\Theta}}
\def \Tr {{\rm Tr}}
\def\CA{{\cal A}}
\def\CC{{\cal C}}
\def\CD{{\cal D}}
\def\CE{{\cal E}}
\def\CF{{\cal F}}
\def\CH{{\cal H}}
\def\CJ{{\cal J}}
\def\CK{{\cal K}}
\def\CL{{\cal L}}
\def\CM{{\cal M}}
\def\CN{{\cal N}}
\def\CO{{\cal O}}
\def\CP{{\cal P}}
\def\CQ{{\cal Q}}
\def\CR{{\cal R}}
\def\CS{{\cal S}}
\def\CT{{\cal T}}
\def\CV{{\cal V}}
\def\CW{{\cal W}}
\def\CY{{\cal Y}}
\def\BC{\mathbb{C}}
\def\BR{\mathbb{R}}
\def\BZ{\mathbb{Z}}
\def\sA{\mathscr{A}}
\def\sB{\mathscr{B}}
\def\sF{\mathscr{F}}
\def\sG{\mathscr{G}}
\def\sH{\mathscr{H}}
\def\sJ{\mathscr{J}}
\def\sL{\mathscr{L}}
\def\sM{\mathscr{M}}
\def\sN{\mathscr{N}}
\def\sO{\mathscr{O}}
\def\sP{\mathscr{P}}
\def\sR{\mathscr{R}}
\def\sQ{\mathscr{Q}}
\def\sS{\mathscr{S}}
\def\sX{\mathscr{X}}

\def\slz{SL(2,Z)}
\def\slr{$SL(2,R)\times SL(2,R)$ }
\def\ads{${AdS}_5\times {S}^5$ }
\def\adst{${AdS}_3$ }
\def\sun{SU(N)}
\def\ad#1#2{{\frac \delta {\delta\sigma^{#1}} (#2)}}
\def\bqf{\bar Q_{\bar f}}
\def\nf{N_f}
\def\sunf{SU(N_f)}

\def\dcirc{{^\circ_\circ}}

\author{Morgan H. Lynch}
\email{mhlynch@uwm.edu}
\affiliation{Leonard E. Parker Center for Gravitation, Cosmology and Astrophysics, Department of Physics, University of Wisconsin-Milwaukee,
P.O.Box 413, Milwaukee, Wisconsin 53201, USA}

\title{Acceleration-induced scalar field transitions of $n$-particle multiplicity}
\date{\today}

\begin{abstract}
In this paper we calculate the effect of acceleration on the decay and excitation rates of scalar fields into a final state of arbitrary multiplicity. The analysis is carried out using standard field operators as well as an Unruh-DeWitt detector. Using the equivalence of the two methods, we show how to correctly set up the computation and interpret the results in terms of the particle content of the initial and final state Rindler and Minkowski spacetimes. We find the dominant transition pathway, and thus final state multiplicity, is acceleration dependent. The formalisms developed are then used to analyze the electron and muon system. We compute the transition rates and lifetimes for accelerated electrons and muons as well as the branching fractions for muon decay. 
\end{abstract}

\pacs{04.62.+v, 13.20.-v}

\maketitle

\section{Introduction}
Since the discoveries of Parker [1], Hawking [2], and Unruh [3], namely cosmological particle creation, black hole evaporation, and accelerated radiation, respectively, a general notion has emerged that the particle content of spacetime is an observer-dependent quantity. For example, with the Unruh effect an observer undergoing uniform acceleration $a$ will find the Minkowski vacuum state to be a thermalized bath of particles at temperature $t = a/2\pi$. Directly measuring this, or related phenomena, has remained outside the reach of our current experimental capabilities. Indirect measurements, such as the acceleration-dependent lifetime of particles, could provide a better avenue for verifying these effects. Muller [4] first calculated how acceleration affects the decay rates of muons, pions, and protons using scalar fields. A more detailed calculation of the accelerated decay of protons and neutrons, and related processes, using fermions coupled to semiclassical vector currents was carried out by Matsas and Vanzella [5-7]. The weak decay processes that have been considered so far have final states containing only two or three particles. By generalizing the formalism to arbitrary final state multiplicities we are able to model all decay processes regardless of the number of daughter products and gain insight into how the branching fractions of the various decay chains change with acceleration. The scalar field formalisms developed can be applied to a wide range of weak decay processes including the previously analyzed cases of proton, neutron, pion, and muon decays. A comprehensive analysis of how the branching fraction of these processes evolve under acceleration has yet to be carried out. This paper carries out the branching fraction analysis for the muon and also gives a first estimate for the lifetime of an accelerated electron using a scalar field approximation.

In this paper, Sec. II focuses on calculating the transition rates and lifetimes for an accelerated particle to decay into $n_{M}$ massless Minkowski particles. The calculation is carried out using standard field operators operating on the fock states of their respective spacetimes. We derive the Wightman functions and then evaluate them along the trajectory of the accelerated particle. This formalism is effectively a lab frame calculation of the transition rates. In Sec. III, we use an Unruh-DeWitt detector to model the inclusion of a massive final state and calculate the transition rates and lifetimes for an accelerated two-level system to undergo a transition with the simultaneous emission of $n_{M}$ massless particles into Minkowski space. We also insert the trajectory prior to calculating the Wightman functions. The subsequent calculations give insight into the physics in the rest frame of the detector and are effectively a proper frame calculation of the transition rates. Section. IV deals with the comparison of the first two methods of calculation. We show how to calculate the transition rate for an initially accelerated particle to decay into $n_{R}$ particles of arbitrary energy into Rindler space and $n_{M}$ massless particles into Minkowski space. In Sec. V we apply the formalisms to model the accelerated weak decay of muons and the accelerated excitation of electrons back into muons. The acceleration-dependent branching fractions of muon decay are also included in the analysis. Section. VI summarizes the conclusions of the manuscript. We use natural units $\hbar = c = k_{B} = 1$ throughout. 

\section{Method of Field Operators}

In this section we determine the probability per unit time that a massive scalar particle will decay into $n_{M}$ massless scalar particles using the method of field operators. Denoting the massive initial state by $\Psi$ and the massless final states by $\phi_{i}$, the process we are concerned with is given by

\bqe
\Psi \rightarrow_{a} \phi_{1}\phi_{2}\phi_{3}\cdots\phi_{n_{M}}.
\eqe

It should be noted that there may be symmetry factors associated with the final state products if there are more than one of the same particle species in the final state. For the current considerations we ignore any symmetry factors which may arise since we will have an arbitrary coupling constant which may be rescaled to take into account any degeneracy, statistical, or color factors. In order to describe this decay process, we work in the interaction picture and consider the following action:

\bqe
\hat{S}_{I} = \int d^{4}x\sqrt{-g} \sqrt{\frac{2}{\sigma \kappa}}G \hat{\Psi} \prod_{\ell = 1}^{n_{M}}\hat{\phi}_{\ell}.
\eqe

The coupling constant $G$ will be determined by the specific interaction and, for the eventual concern of this paper, will be related to the Fermi coupling $G_{f}$. The additional factor of $\sqrt{\frac{2}{\sigma \kappa}}$ will be used for the later convenience of absorbing the Jacobian of a proper time reparametrization and normalization constant. Note that we are modeling decay processes at tree level and provided the energy scale, i.e. the proper acceleration, remains below the $W^{\pm}$ and $Z$ boson masses we need not worry about the nonrenormalizability of this effective Fermi interaction. All fields under consideration are assumed to be real and thus so is the interaction action. Note, all interactions, fields, trajectories, and thus the transition rate will eventually be evaluated in the Rindler coordinate chart. The probability amplitude for the acceleration induced decay of our massive initial state into $n_{M}$ massless particles is given by

\bqa
\mathcal{A} = \bra{\prod_{m = 1}^{n_{M}}\mathbf{k}_{m}} \otimes \bra{0} \hat{S}_{I}\ket{\Psi_{i}} \otimes \ket{0}.
\eqa

That is, the initial fock state $\ket{\Psi_{i}}$ of our massive field $\Psi$ decays into the $n_{M}$-particle momentum eigenstate $\ket{\prod_{i = 1}^{n_{M}}\mathbf{k}_{i}}$ of our massless fields $\phi_{i}$ under the influence of the interaction $\hat{S}_{I}$. Note we have used the shorthand notation $\ket{\prod_{i = 1}^{n_{M}}\mathbf{k}_{i}} = \ket{\mathbf{k}_{1},\mathbf{k}_{2},\ldots, \mathbf{k}_{n_{M}}}$ to denote our final state. Defining $\prod_{j = 1}^{n_{M}}d^{3}k_{j} = D_{n_{M}}^3 k$, we can set up the differential probability, i.e. the magnitude squared of the probability amplitude per unit final state momenta, via

\bqa
\frac{d\mathcal{P}}{D_{n_{M}}^3 k} &=& |\mathcal{A}|^{2} \nonumber \\ 
&=&  \left| \bra{\prod_{m = 1}^{n_{M}}\mathbf{k}_{m}} \otimes \bra{0} \hat{S}_{I}\ket{\Psi_{i}} \otimes \ket{0} \right|^{2}\nonumber \\ 
&=&  G^2 \frac{2}{\sigma \kappa}  \int d^{4}x\sqrt{-g} \int d^{4}x'\sqrt{-g'}\left| \bra{\prod_{m = 1}^{n_{M}}\mathbf{k}_{m}} \otimes \bra{0}  \hat{\Psi}(x) \prod_{\ell = 1}^{n_{M}}\hat{\phi}_{\ell}(x)\ket{\Psi_{i}} \otimes \ket{0} \right|^{2} \non \\
&=&  G^2 \frac{2}{\sigma \kappa}  \int d^{4}x\sqrt{-g} \int d^{4}x'\sqrt{-g'}\left| \bra{0}  \hat{\Psi}(x)  \ket{\Psi_{i}} \right|^{2}\left| \bra{\prod_{m = 1}^{n_{M}}\mathbf{k}_{m}}\prod_{\ell = 1}^{n_{M}}\hat{\phi}_{\ell}(x)\ket{0} \right|^{2}.
\eqa

The above inner product containing our massless fields $\phi_{\ell}$, its complex conjugate, and the product of momentum integrations in Eq. (4) allow us to factor out $n_{M}$ complete sets of momentum eigenstates, e.g. $\int d^{3}k \ket{k}\bra{k} =1$. The total transition probability is then given by

\bqa
\mathcal{P} &=&  G^2 \frac{2}{\sigma \kappa}  \int d^{4}x\sqrt{-g} \int d^{4}x'\sqrt{-g'}\left| \bra{0}  \hat{\Psi}(x)  \ket{\Psi_{i}} \right|^{2} \prod_{j = 1}^{n_{M}}\int d^{3}k_{j} \left| \bra{\prod_{m = 1}^{n_{M}}\mathbf{k}_{m}}\prod_{\ell = 1}^{n_{M}}\hat{\phi}_{\ell}(x)\ket{0} \right|^{2} \non \\
&=&  G^2 \frac{2}{\sigma \kappa}  \int d^{4}x\sqrt{-g} \int d^{4}x'\sqrt{-g'}\left| \bra{0}  \hat{\Psi}(x)  \ket{\Psi_{i}} \right|^{2} \prod_{\ell = 1}^{n_{M}} \bra{0}\hat{\phi}_{\ell}(x')\hat{\phi}_{\ell}(x)\ket{0}.
\eqa

In examining the above equation, it serves to recall the expression $\bra{0}\hat{\Psi}(x)\ket{\Psi_{i}}$ selects the positive frequency mode function $u_{k}(x,\tau)$ of the initial state $\Psi$. These positive frequency mode functions are eigenfunctions of the Rindler coordinate proper time $\tau$ such that $\partial_{\tau}u_{k} = -i\omega u_{k}$. In the accelerated frame this particle is at rest and its energy is only the rest mass $m$. Letting $f_{\Psi_{i}}(x)$ denote the spatial variation of the particle, we find

\bqa
\bra{0}\hat{\Psi}(x)\ket{\Psi_{i}} &=& \bra{0}\int d^3k' [\hat{a}_{k'}u_{k'}(x) + h.c]\ket{\Psi_{i}} \nonumber \\
&=& \int d^3k' \delta(k'-k)u_{k'}(x) \nonumber \\
&=& u_{k}(x) \nonumber \\
&=& f_{\Psi_{i}}[x(\tau)]e^{-im\tau}.
\eqa  

Furthermore, each of the two-point functions $\bra{0}\hat{\phi}_{\ell}(x')\hat{\phi}_{\ell}(x)\ket{0}$ in Eq. (5) characterizes the probability amplitude for a field quanta to be created at the spacetime point $x$ and propagate within the lightcone to the spacetime point $x'$. If $t'>t$ then the particle is traveling forward through time and has a postive frequency. This defines the appropriately named positive frequency Wightman function denoted $G^{+}(x',x)$. Similarly if $t>t'$ then this defines the negative frequency Wightman function, denoted $G^{-}(x',x)$, and describes a particle of negative frequency propagating backwards through time. The time ordered sum of the positive and negative frequency Wightman functions make up the more common Feynman propagator [8]. Denoting the general two point function $G^{\pm}(x',x)$, our probability can now be simplified to the following form:

\bqa
\mathcal{P} &=& G^2 \frac{2}{\sigma \kappa}  \int d^{4}x\sqrt{-g} \int d^{4}x'\sqrt{-g'}\left| \bra{0}  \hat{\Psi}(x)  \ket{\Psi_{i}} \right|^{2} \prod_{\ell = 1}^{n_{M}} \bra{0}\hat{\phi}_{\ell}(x')\hat{\phi}_{\ell}(x)\ket{0} \nonumber \\
&=& G^2 \frac{2}{\sigma \kappa} \int d^{4}x\sqrt{-g} \int d^{4}x'\sqrt{-g'} f_{\Psi_{i}}(x)f^{\ast}_{\Psi_{i}}(x')e^{im(\tau'-\tau)} [ G^{\pm}(x',x)]^{n_{M}}.
\eqa

The Wightman functions for the massless scalar field can be evaluated analytically by inserting the canonically normalized mode decomposition of our field operator $\hat{\phi} = \int \frac{ d^{3}k}{(2 \pi)^{3/2}\sqrt{2\omega}} [ \hat{a}_{\mathbf{k}}e^{i(\mathbf{k}\cdot \mathbf{x} -\omega t)}  + \hat{a}_{\mathbf{k}}^{\dagger}e^{-i(\mathbf{k}\cdot \mathbf{x} -\omega t)}]$. Thus,

\bqa
G^{\pm}(x',x) &=& \bra{0_{\ell}}\hat{\phi}_{\ell}(x')\hat{\phi}_{\ell}(x)\ket{0_{\ell}} \nonumber \\
&=& \frac{1}{2(2 \pi)^{3}}\iint\frac{ d^{3}k'd^{3}k}{\sqrt{\omega'\omega}}\bra{0_{\ell}}\lbk \hat{a}_{\mathbf{k'}}e^{i(\mathbf{k'}\cdot \mathbf{x'} -\omega' t')}  + \hat{a}_{\mathbf{k'}}^{\dagger}e^{-i(\mathbf{k'}\cdot \mathbf{x'} -\omega' t')}\rbk\lbk \hat{a}_{\mathbf{k}}e^{i(\mathbf{k}\cdot \mathbf{x} -\omega t)}  + \hat{a}_{\mathbf{k}}^{\dagger}e^{-i(\mathbf{k}\cdot \mathbf{x} -\omega t)}\rbk\ket{0_{\ell}} \nonumber \\
&=& \frac{1}{2(2 \pi)^{3}}\iint\frac{ d^{3}k'd^{3}k}{\sqrt{\omega'\omega}}\bra{0_{\ell}} \hat{a}_{\mathbf{k'}}\hat{a}_{\mathbf{k}}^{\dagger}e^{i(\mathbf{k'}\cdot \mathbf{x'} -\mathbf{k}\cdot \mathbf{x} -\omega' t'+\omega t)}\ket{0_{\ell}} \nonumber \\
&=& \frac{1}{2(2 \pi)^{3}}\iint\frac{ d^{3}k'd^{3}k}{\sqrt{\omega'\omega}}e^{i(\mathbf{k'}\cdot \mathbf{x'} -\mathbf{k}\cdot \mathbf{x} -\omega' t'+\omega t)}\delta(\mathbf{k'}-\mathbf{k}) \nonumber \\
&=& \frac{1}{2(2 \pi)^{3}}\int\frac{d^{3}k}{\omega}e^{i(\mathbf{k}\cdot \Delta\mathbf{x} -\omega \Delta t)}.
\eqa   

To facilitate the resultant integral we move into momentum space spherical coordinates and rotate until our momentum is aligned along the $z$ axis. Recall that in the massless limit $\omega = k$ the integration simplifies further to

\bqa
G^{\pm}(x',x) &=& \frac{1}{2(2 \pi)^{3}}\int\frac{d^{3}k}{\omega}e^{i(\mathbf{k}\cdot \Delta\mathbf{x} -\omega \Delta t)} \nonumber \\
&=& \frac{1}{2(2 \pi)^{3}}\int_{0}^{\infty}\int_{0}^{\pi}\int_{0}^{2\pi} dkd\theta d\phi \;k \sin{\theta}e^{i(k\Delta x \cos{\theta} -k \Delta t)} \nonumber \\
&=& \frac{1}{2(2 \pi)^{2}}\int_{0}^{\infty}\int_{-1}^{1} dkd(\cos{\theta}) \;k e^{i(k\Delta x \cos{\theta} -k \Delta t)} \nonumber \\
&=& \frac{1}{2(2 \pi)^{2}}\frac{i}{\Delta x} \int_{0}^{\infty} dk \lbk e^{-ik(\Delta x + \Delta t)} - e^{-ik(-\Delta x + \Delta t)} \rbk.
\eqa

In order for the above integration to be well defined we must damp the oscillation at infinity via the introduction of a complex regulator to our time interval, e.g. $\Delta t \rightarrow \Delta t - i\epsilon$ with $\epsilon >0$. Hence,

\bqa
G^{\pm}(x',x) &=& \frac{1}{2(2 \pi)^{2}}\frac{i}{\Delta x} \int_{0}^{\infty} dk \lbk e^{-ik(\Delta x + \Delta t)} - e^{-ik(-\Delta x + \Delta t)} \rbk \nonumber \\
&=& \frac{1}{2(2 \pi)^{2}}\frac{i}{\Delta x} \int_{0}^{\infty} dk \lbk e^{-ik(\Delta x + (\Delta t-i\epsilon))} - e^{-ik(-\Delta x + (\Delta t -i\epsilon))} \rbk \nonumber \\
&=&\frac{1}{2(2 \pi)^{2}}\frac{i}{\Delta x}\lbk \frac{1}{i(\Delta x + (\Delta t-i\epsilon))} - \frac{1}{i(-\Delta x + (\Delta t-i\epsilon))}  \rbk \nonumber \\
&=&\frac{1}{(2 \pi)^{2}} \frac{1}{\Delta x^{2} - (\Delta t-i\epsilon)^{2}}.
\eqa

Having determined the functional form of our massless Wightman function we return to the integrations over the spatial coordinates in our decay probability, Eq. (7). These can be dealt with by examining the covariant 4-volume element of Rindler space. The proper coordinates [9] $(\tau,\xi,\mathbf{x}_{\bot})$ seen by a particle undergoing uniform proper acceleration $a$ along the $z$ axis are given by

\bqa
\tau(t,z) = \frac{1}{2a}\ln{\frac{z+t}{z-t}} \nonumber \\ 
\xi(t,z) = -\frac{1}{a}+\sqrt{z^{2}-t^{2}}.
\eqa

The perpendicular coordinates $\mathbf{x}_{\perp}$ do not change in Rindler space. Note, the coordinate $\xi$ parametrizes distances seen by the accelerated observer along the axis of acceleration and the point $\xi = 0$ labels the origin of this axis and is defined to be the location of the uniformly accelerated particle. For an inertial observer, this point will then characterize the trajectory of the accelerated particle. In this coordinate chart, the metric takes the form

\bqe
ds^{2} = (1+a\xi)^{2}d\tau^2 - d\xi^2 -d\mathbf{x}_{\perp}^{2}.
\eqe

The corresponding metric determinant of this spacetime used to covariantly scale our 4-volume of integration is $|g|=1+a\xi$. Inverting our proper coordinate chart, Eq. (11), and translating until $\xi = 0$ and $\mathbf{x}_{\bot} = 0$ yields the trajectory of our particle,

\bqa
t &=& \frac{1}{a}\sinh{a\tau} \nonumber \\
z &=& \frac{1}{a}\cosh{a\tau} \nonumber \\
\mathbf{x}_{\perp} &=& 0.
\eqa 

It should be noted that under this trajectory our Wightman function, Eq. (10), depends only on the proper time $\tau$ and is therefore not affected by the spatial integrations. Returning to the decay probability, Eq. (7), we can handle the spatial components of the integration via

\bqa
\mathcal{P} &=& G^2 \frac{2}{\sigma \kappa} \int d^{4}x\sqrt{-g} \int d^{4}x'\sqrt{-g'} f_{\Psi}(x)f^{\ast}_{\Psi}(x')e^{im(\tau'-\tau)} [ G^{\pm}(x',x)]^{n_{M}} \nonumber \\
&=& G^2 \frac{2}{\sigma \kappa} \int d^{3}x\sqrt{1+a\xi} \int d^{3}x'\sqrt{1+a\xi'} f_{\Psi}(x)f^{\ast}_{\Psi}(x') \iint d\tau d\tau' e^{im(\tau'-\tau)} [ G^{\pm}(x',x)]^{n_{M}} \nonumber \\
&=& G^2 \frac{2}{\sigma \kappa} \kappa \iint d\tau  d\tau' e^{im(\tau'-\tau)} \lbk G^{\pm}(x',x)\rbk^{n_{M}} \nonumber \\
&=& G^2 \frac{2}{\sigma } \mathcal{F}_{n_{M}}(m).
\eqa 

The mode functions have the form $f_{\Psi}(x) \sim K_{i\omega/a}(\frac{m}{a}e^{a\xi}) g(\mathbf{x}_{\perp})$ where $g(\mathbf{x}_{\perp})$ is an envelope function or wave packet describing the spatial distribution of our accelerated field in the directions perpendicular to the acceleration. With the mode functions properly normalized [10], the expression $\kappa = \left| \int d^{3}x\sqrt{1+a\xi}f_{\Psi}(x)\right|^{2}$ will be of order unity. We see the probability of an acceleration induced transition is then given by the Fourier transform of the product of the $n_{M}$ final state Wightman functions. This is known as the response function $\mathcal{F}_{n_{M}}(m)$. The effective coupling constant $G^{2}$ for the process being considered will be determined by taking the limit $a\rightarrow 0$ and matching the coefficient to the known inertial decay process. Note this compact form of the transition probability is valid for a more general class of trajectories provided their parametrization only depends on the proper time. Using the trajectories from Eq. (13), the coordinate transformations $u = \frac{\tau'-\tau}{\rho}$ and $s = \frac{\tau'+\tau}{\sigma}$, and the inversion $\tau'=\frac{\rho u + \sigma s }{2}$ and $\tau =\frac{\sigma s - \rho u}{2}$ with arbitrary $\sigma$ and $\rho$, we find the explicit form of the spacetime intervals in the massless Wightman function to be

\bqa
\Delta x^{2} - (\Delta t-i\epsilon)^{2} &=& \frac{1}{a^2}\left\lbrace [\cosh{\lb a\tau'\rb}-\cosh{\lb a \tau \rb}]^2 -[\sinh{\lb a\tau' \rb}-\sinh{\lb a \tau \rb}-i\epsilon]^2\right\rbrace \nonumber \\
&=& \frac{1}{a^2}\left\lbrace [2\sinh{\lb\frac{a\rho u}{2}\rb}\sinh{\lb\frac{a \sigma s}{2}\rb}]^2 -[2\sinh{\lb\frac{a\rho u}{2}\rb}\cosh{\lb\frac{a\sigma s}{2}\rb}-i\epsilon]^2\right\rbrace \nonumber \\
&=& \frac{1}{a^2} [4\sinh^{2}{\lb\frac{a\rho u}{2}\rb}\sinh^{2}{\lb\frac{a\sigma s}{2}\rb} -4\sinh^{2}{\lb\frac{a\rho u}{2}\rb}\cosh^{2}{\lb\frac{a\sigma s}{2}\rb}+4i\epsilon\sinh{\lb\frac{a\rho u}{2}\rb}\cosh{\lb\frac{a\sigma s}{2}\rb}] \nonumber \\
&=& \frac{1}{a^2} [4\sinh^{2}{\lb\frac{a\rho u}{2}\rb}\sinh^{2}{\lb\frac{a\sigma s}{2}\rb} -4\sinh^{2}{\lb\frac{a\rho u}{2}\rb}\cosh^{2}{\lb\frac{a\sigma s}{2}\rb}+8i\epsilon\sinh{\lb\frac{a\rho u}{2}\rb}\cosh{\lb\frac{a\rho u}{2}\rb}] \nonumber \\
&=& \frac{-4}{a^2}\sinh^{2}{\lb\frac{a\rho u}{2}-i\epsilon\rb}.
\eqa

Note we have rescaled $\epsilon$ by the positive definite factor $2\cosh{\lb\frac{a\rho u}{2}\rb}/\cosh{\lb\frac{a\sigma s}{2}\rb}$ and used the Taylor expansion of $\sinh^2(x-i\epsilon)$ to combine the arguments. Thus we obtain

\bqe
G^{\pm}(x',x) = -\frac{1}{(2 \pi)^{2}} \frac{a^{2}}{4\sinh^{2}{\lb\frac{a\rho u}{2}-i\epsilon\rb}}.
\eqe

In changing the proper time integration variables we pick up the Jacobian $\frac{\sigma \rho}{2}$ and our transition probability induced by the uniformly accelerated trajectory then becomes

\bqa
\mathcal{P} &=& G^2 \frac{2}{\sigma}  \mathcal{F}_{n_{M}}(m) \nonumber \\
&=& G^2 \frac{2}{\sigma} \iint d\tau  d\tau' e^{im(\tau'-\tau)} \lbk G^{\pm}(x',x)\rbk^{n_{M}} \nonumber \\
&=& G^2 \frac{2}{\sigma}  \iint d\tau d\tau' e^{im(\tau'-\tau)} \lbk \frac{1}{(2 \pi)^{2}} \frac{1}{\Delta x^{2} - (\Delta t-i\epsilon)^{2}}\rbk^{n_{M}} \nonumber \\
&=& G^2 \frac{2}{\sigma}  \iint d\tau d\tau' e^{im(\tau'-\tau)} \lbk -\frac{1}{(2 \pi)^{2}} \frac{a^{2}}{4\sinh^{2}{\lb\frac{a\rho u}{2}-i\epsilon\rb}}\rbk^{n_{M}} \nonumber \\
&=& G^2  (-1)^{n_{M}} \rho \lb \frac{a}{4 \pi}  \rb^{2n_{M}} \iint ds du \; \frac{e^{im\rho u}}{[\sinh{\lb\frac{a\rho u}{2}-i\epsilon\rb}]^{2n_{M}}}.
\eqa

By dividing out the infinite proper time interval $\int ds$ we obtain the probability of transition per unit proper time $\Gamma_{n_{M}}(m,a) = \frac{\mathcal{P}}{\Delta s}$. After rescaling $u \rightarrow \rho u$ we see that the result is independent of the parametrization of $u$. The parametrization of $s$ yielded a factor of $\frac{\sigma}{2}$ which we absorbed by the initial rescaling of our coupling constant. The probability per unit time is thus given by 

\bqe
\Gamma_{n_{M}}(m,a) = G^2  \lb \frac{ia}{4 \pi}  \rb^{2n_{M}} \int du \; \frac{e^{imu}}{[\sinh{\lb\frac{a u}{2}-i\epsilon\rb}]^{2n_{M}}}.
\eqe

Focusing on the integration, we note that in the absence of the $i\epsilon$ prescription there will be poles of order $2n_{M}$ when $u = 2\alpha\pi i/a$ with $\alpha$ being any integer. To integrate over the real axis in the presence of the pole at $u = 0$ we will close our contour in the upper half plane to damp the oscillation at infinity. In doing so we also pick up the additional tower of poles along the imaginary axis. Furthermore, with the negative $i\epsilon$ prescription we will also capture the pole at $\alpha=0$. We will now remove the regulator $\epsilon \rightarrow 0$ now that we understand the appropriate pole structure. The integrand can be cast into a simpler form via the change of variables $w = e^{au}$. Hence,

\bqa
\int du \; \frac{e^{imu}}{[\sinh{\lb\frac{a u}{2}\rb}]^{2n_{M}}} &=& 2^{2n_{M}}\int_{-\infty}^{\infty} du \; \frac{e^{imu}}{\lbk e^{\frac{au}{2}}-e^{-\frac{au}{2}}\rbk^{2n_{M}}} \nonumber \\
&=& 2^{2n_{M}}\int_{-\infty}^{\infty} du \; \frac{e^{imu+aun_{M}}}{\lbk e^{au}-1 \rbk^{2n_{M}}} \nonumber \\
&=& \frac{2^{2n_{M}}}{a}\int_{0}^{\infty} dw \; \frac{w^{im/a+n_{M}-1}}{\lbk w-1 \rbk^{2n_{M}}}.
\eqa

We see that there are poles when $w=1$, i.e. $w = e^{i2\pi \alpha}$ where we keep the integer $\alpha\geq0$. Evaluation of this integral may be accomplished via the residue theorem.  Thus

\bqa
\frac{2^{2n_{M}}}{a}\int_{0}^{\infty} dw \; \frac{w^{im/a+n_{M}-1}}{\lbk w-1 \rbk^{2n_{M}}} &=& \frac{2^{2n_{M}}}{a} \frac{2 \pi i}{(2n_{M}-1)!}\sum_{\alpha = 0}^{\infty}\frac{d^{2n_{M}-1}}{dw^{2n_{M}-1}}\lbk\lbk w-1 \rbk^{2n_{M}}\frac{w^{im/a+n_{M}-1}}{\lbk w-1 \rbk^{2n_{M}}}\rbk_{w = e^{i2\pi \alpha}} \nonumber \\
&=& \frac{2^{2n_{M}}}{a} \frac{2 \pi i}{(2n_{M}-1)!}\sum_{\alpha = 0}^{\infty}\lbk\frac{w^{im/a-n_{M}}\Gamma(im/a+n_{M})}{\Gamma(im/a+1-n_{M})}\rbk_{w = e^{i2\pi \alpha}} \nonumber \\
&=& \frac{2^{2n_{M}}}{a} \frac{2 \pi i}{(2n_{M}-1)!}\frac{\Gamma(im/a+n_{M})}{\Gamma(im/a+1-n_{M})}\sum_{\alpha = 0}^{\infty} e^{-2\pi\frac{m}{a}\alpha-2\pi in\alpha} \nonumber \\
&=& \frac{2^{2n_{M}}}{a} \frac{2 \pi i}{(2n_{M}-1)!}\frac{\Gamma(im/a+n_{M})}{\Gamma(im/a+1-n_{M})}\frac{1}{1-e^{-2\pi m/a}}. 
\eqa 

The presence of the factor of $[1-e^{-2\pi m/a}]^{-1}$ is indicative of the thermal nature of the vacuum associated with the Unruh effect. From our total rate, Eq. (18), for a uniformly accelerated particle of mass $m$ to decay into $n_{M}$ massless particles under the influence of a uniform acceleration is then found to be

\bqe
\Gamma_{n_{M}}(m,a) = G^2 \lb \frac{ia}{2 \pi}  \rb^{2n_{M}} \frac{1}{a} \frac{2 \pi i}{(2n_{M}-1)!}\frac{\Gamma(im/a+n_{M})}{\Gamma(im/a+1-n_{M})}\frac{1}{1-e^{-2\pi m/a}}.
\eqe

We can normalize the above expression by defining $\tilde{\Gamma}=\Gamma/\Gamma_{0}$, with $\Gamma_{0} = G^{2}$, to better analyze the normalized decay rate for an arbitrary $n_{M}$ particle multiplicity final state. The normalized decay rates $\tilde{\Gamma}_{n_{M}}$ for the first few integer values of $n_{M}$ are given by

\bqa
\tilde{\Gamma}_{1}(m,a) &=& \frac{m}{2 \pi} \frac{1}{1-e^{-2\pi m/a}} \non \\
\tilde{\Gamma}_{2}(m,a) &=& \frac{m^{3}}{48 \pi^{3}} \frac{1+\lb \frac{a}{m}  \rb^{2}}{1-e^{-2\pi m/a}} \non \\
\tilde{\Gamma}_{3}(m,a) &=& \frac{m^{5}}{3840 \pi^{5}} \frac{1+ 5 \lb \frac{a}{m}  \rb^{2} + 4 \lb \frac{a}{m}  \rb^{4}}{1-e^{-2\pi m/a}} \non \\
\tilde{\Gamma}_{4}(m,a) &=& \frac{m^{7}}{645120 \pi^{7}} \frac{1+ 14 \lb \frac{a}{m}  \rb^{2} + 49 \lb \frac{a}{m}  \rb^{4} + 36 \lb \frac{a}{m}  \rb^{6}}{1-e^{-2\pi m/a}} \non \\
\tilde{\Gamma}_{5}(m,a) &=& \frac{m^{9}}{185794560 \pi^{9}} \frac{1+ 30 \lb \frac{a}{m}  \rb^{2} + 273 \lb \frac{a}{m}  \rb^{4} + 820 \lb \frac{a}{m}  \rb^{6} + 576 \lb \frac{a}{m}  \rb^{8} }{1-e^{-2\pi m/a}}.
\eqa

Below, in Figs. 1 and 2, we plot both the normalized decay rates and lifetimes $\tilde{\tau} = 1/\tilde{\Gamma}$ for a particle of mass $m = 1$ to decay into $n_{M}$ massless particle states as a function of the proper acceleration. It is clear from both Eq. (22) and the plots below that there exists a crossover scale of acceleration where the accelerated particle will preferentially choose the decay chain with the most final state products. This implies that an inertially decaying particle chooses the decay chain which contains the least allowable amount of end products and by imparting a sufficiently high acceleration on an unstable particle it will chose the decay chain which contains the most allowable final state products. 

\begin{figure}[H]
\centering  
\includegraphics[,scale=.6]{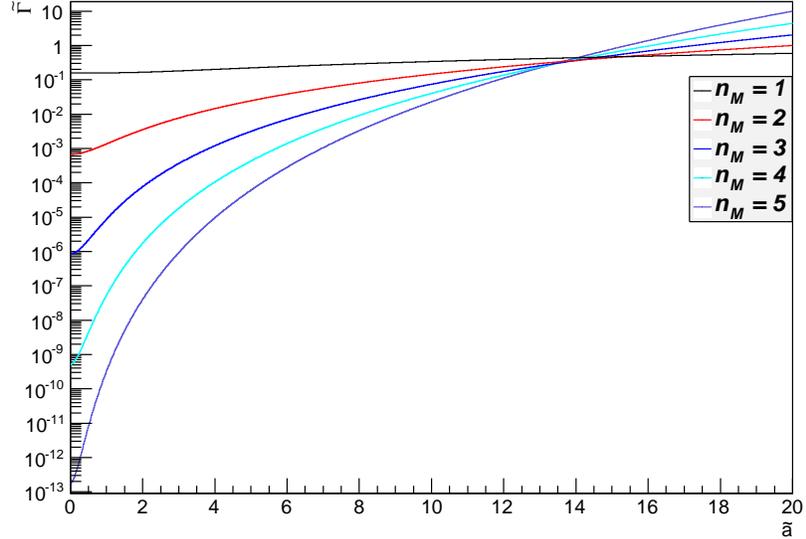}
\caption{The normalized decay rates, Eq. (22), with $\tilde{a} = a/m$ and $m=1$.}
\end{figure}  

\begin{figure}[H]
\centering  
\includegraphics[,scale=.6]{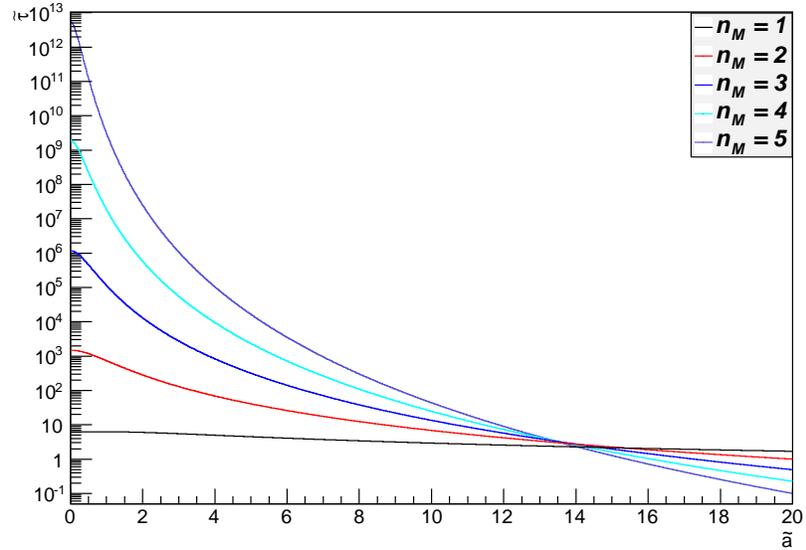}
\caption{The normalized lifetimes, $\tilde{\tau}$, with $\tilde{a} = a/m$ and $m=1$.}
\end{figure}  

The prescription for this method of calculation can be seen by inspecting Eq. (14). In general, for $n_{M}$ final state products, the response function is computed by taking the Fourier transform of the product of each of the Wightman functions of the $n_{M}$ massless final states. The number of final states determines the number of derivatives taken in calculating the residues of Eq. (20), which yields the gamma functions, and thus the number of terms in the decay rate polynomial as can be seen by Eq. (22). In the next section we will analyze the same situation utilizing an Unruh-DeWitt detector. 

\section{The Method of Detectors}

In this section we utilize the formalism of Unruh-DeWitt detectors; see Ref. [3,11]. As such, we form a two-level system consisting of two particles of arbitrary mass and determine the associated decay and excitation rates, accompanied by the simultaneous emission of $n_{M}$ massless particles, of the system under uniform acceleration. These processes are illustrated schematically as

\bqe
\Psi_{1} \rightarrow_{a} \Psi_{2}\phi_{1}\phi_{2}\cdots\phi_{n_{M}}.
\eqe

The utility of this method is that it allows the inclusion of a massive final state in a rather uncomplicated fashion and, more importantly, allows for a description of acceleration induced excitation rather than just decay. To accomplish this, we will now promote the massive scalar fields $\Psi_{i}$ to a two-level system, e.g. an Unruh-DeWitt detector. These fields, and their transitions, will now be characterized by the time evolved monopole moment operator,

\bqe
\hat{m}(\tau) = e^{i\hat{H}\tau}\hat{m}_{0}e^{-i\hat{H}\tau}.
\eqe

The monopole moment operator $\hat{m}_{0}$ is assumed to be Hermitian. The operator $\hat{H}$ denotes the detectors or fields proper Hamiltonian with the property

\bqe
\hat{H}\ket{\Psi_{i}} = m_{i}\ket{\Psi_{i}}, \; i = 1,2
\eqe

since, in the proper frame, the total energy will be that of the rest mass of our field $m_{i}$. Utilizing this formalism, we define the interaction action as

\bqe
\hat{S}_{I} = \int d\tau \, \sqrt{\frac{2}{\sigma}}\hat{m}(\tau)\prod_{\ell = 1}^{n_{M}}\hat{\phi}_{\ell}.
\eqe  

Again we have pulled out the additional factor of $\sqrt{\frac{2}{\sigma}}$ to absorb the Jacobian of a proper time reparametrization. Furthermore, this action is only integrated over the detector proper time and not the full spatial extent of the accelerated field as in the previous section, Eq. (2), since we are considering the fields as a time-dependent two-level system with no spatial extent. In calculating matrix elements of the form $\bra{\Psi_{f}}\hat{m}(\tau)\ket{\Psi_{i}}$ we define the effective coupling constant to be

\bqe
G = \bra{\Psi_{f}}\hat{m}_{0}\ket{\Psi_{i}}.
\eqe

It is this effective coupling constant that encodes the physical characteristics of the particular transition under consideration. The probability amplitude for the process induced by the interaction, Eq. (26), is given by

\bqe
\mathcal{A} = \bra{\prod_{j = 1}^{n_{M}}\mathbf{k}_{j}}\otimes \bra{\Psi_{f}} \hat{S}_{I}\ket{\Psi_{i}} \otimes \ket{0}.
\eqe

We again use the same notation for our Fock states and accommodate any complications due to the statistics or degeneracies of the final state products by rescaling our effective coupling. Utilizing the shorthand notation $\prod_{j = 1}^{n_{M}} d^{3}k_{j} = D^{3}_{n_{M}}k$, the differential probability for the two-level system to undergo a transition and be accompanied by the emission of $n_{M}$ massless particles per unit momentum is given by

\bqa
\frac{d\mathcal{P}}{D^{3}_{n_{M}}k} &=& |\mathcal{A}|^{2} \nonumber \\
&=&  \bra{0} \otimes \bra{\Psi_{i}}   \hat{S}_{I} \ket{\Psi_{f}}\otimes \ket{\prod_{j' = 1}^{n_{M}}\mathbf{k}_{j'}} \bra{\prod_{j = 1}^{n_{M}}\mathbf{k}_{j}}\otimes \bra{\Psi_{f}} \hat{S}_{I}\ket{\Psi_{i}} \otimes \ket{0} \nonumber \\
&=& \frac{2}{\sigma}\iint d\tau'  d\tau \bra{0} \otimes \bra{\Psi_{i}}  \hat{m}(\tau')\prod_{\ell' = 1}^{n_{M}}\hat{\phi}_{\ell'}(x') \ket{\Psi_{f}}\otimes \ket{\prod_{j' = 1}^{n_{M}}\mathbf{k}_{j'}} \bra{\prod_{j = 1}^{n_{M}}\mathbf{k}_{j}}\otimes \bra{\Psi_{f}}  \hat{m}(\tau)\prod_{\ell = 1}^{n_{M}}\hat{\phi}_{\ell}(x)\ket{\Psi_{i}} \otimes \ket{0}.
\eqa  

Operation of the time evolved monopole moment in the relevant inner product and recalling the definition of our effective coupling, Eq. (27), yields

\bqa
\bra{\Psi_{f}}\hat{m}(\tau) \ket{\Psi_{i}} &=& \bra{\Psi_{f}}e^{i\hat{H}\tau}\hat{m}_{0}e^{-i\hat{H}\tau} \ket{\Psi_{i}} \nonumber \\
&=& e^{i(m_{f}-m_{i})\tau}\bra{\Psi_{f}}\hat{m}_{0} \ket{\Psi_{i}} \nonumber \\
&=& Ge^{i\Delta m \tau}.
\eqa

Then our differential probability, Eq. (29), becomes

\bqa
\frac{d\mathcal{P}}{D^{3}_{n_{M}}k} &=& \frac{2}{\sigma}  \iint d\tau' d\tau \;\bra{0} \otimes \bra{\Psi_{i}}  \hat{m}(\tau')\prod_{\ell' = 1}^{n_{M}}\hat{\phi}_{\ell'}(x') \ket{\Psi_{f}}\otimes \ket{\prod_{j' = 1}^{n_{M}}\mathbf{k}_{j'}} \bra{\prod_{j = 1}^{n_{M}}\mathbf{k}_{j}}\otimes \bra{\Psi_{f}}  \hat{m}(\tau)\prod_{\ell = 1}^{n_{M}}\hat{\phi}_{\ell}(x)\ket{\Psi_{i}} \otimes \ket{0} \nonumber \\
&=& G^{2}\frac{2}{\sigma} \iint d\tau' d\tau \;e^{-i\Delta m (\tau' - \tau)} \left| \bra{\prod_{j = 1}^{n_{M}}\mathbf{k}_{j}}  \prod_{\ell = 1}^{n_{M}}\hat{\phi}_{\ell}(x) \ket{0} \right|^{2}.
\eqa

In this section we will endeavor to evaluate the above integral in a different way than in the previous section. Originally we factored out the complete set of momentum eigenstates to yield the Wightman functions. We then showed that each massless Wightman function was, up to a constant, the inverse of the spacetime interval traversed along an arbitrary trajectory. The interval was then evaluated along the hyperbolic trajectory associated with uniform acceleration. Here, we evaluate the inner product without factoring out the complete set of momentum eigenstates. This allows us to insert the hyperbolic trajectory into the resultant mode functions then perform the integrations over momentum. In doing so, we gain insight into the physical properties of the emitted decay products. We also find, as expected, the end result to be identical to that of the previous section. Evaluation of the decay rate using these two different methods lends a greater understanding to the underlying character of these processes. 

Operation on the vacuum with our massless fields in Eq. (31) will yield $n_{M}$ momentum integrals of the negative frequency mode functions over their momentum. Hence the above inner product will reduce to

\bqa
\bra{\prod_{j = 1}^{n_{M}}\mathbf{k}_{j}}  \prod_{\ell = 1}^{n_{M}}\hat{\phi}_{\ell}(x) \ket{0} &=& \bra{\prod_{j = 1}^{n_{M}}\mathbf{k}_{j}}  \prod_{\ell = 1}^{n_{M}} \frac{1}{(2 \pi)^{\frac{3n_{M}}{2}}}\frac{1}{2^{\frac{n_{M}}{2}}}\int \frac{d^{3}k_{\ell}}{\sqrt{ \omega_{\ell}}} \lbk \hat{a}_{\mathbf{k}_{\ell}}^{\dagger} e^{-i(\mathbf{k}_{\ell} \cdot \mathbf{x} -\omega_{k_{\ell}} t)} + h.c. \rbk\ket{0} \nonumber \\
&=& \frac{1}{(2 \pi)^{\frac{3n_{M}}{2}}}\frac{1}{2^{\frac{n_{M}}{2}}}  \prod_{\ell = 1}^{n_{M}} \int \frac{d^{3}k_{\ell}}{\sqrt{ \omega_{\ell}}} e^{-i(\mathbf{k}_{\ell} \cdot \mathbf{x} -\omega_{k_{\ell}} t)} \braket{\prod_{j = 1}^{n_{M}}\mathbf{k}_{j} | \mathbf{k}_{\ell}} \nonumber \\
&=& \frac{1}{(2 \pi)^{\frac{3n_{M}}{2}}}\frac{1}{2^{\frac{n_{M}}{2}}}  \prod_{\ell = 1}^{n_{M}} \int \frac{d^{3}k_{\ell}}{\sqrt{ \omega_{\ell}}} e^{-i(\mathbf{k}_{\ell} \cdot \mathbf{x} -\omega_{k_{\ell}} t)} \prod_{j = 1}^{n_{M}}\delta(\mathbf{k}_{j} - \mathbf{k}_{\ell}) \nonumber \\
&=& \frac{1}{(2 \pi)^{\frac{3n_{M}}{2}}}\frac{1}{2^{\frac{n_{M}}{2}}}   \frac{e^{-i\sum_{j = 1}^{n_{M}}(\mathbf{k}_{j} \cdot \mathbf{x} -\omega_{k_{j}} t)}}{\sqrt{ \prod_{j' = 1}^{n_{M}}  \omega_{j'}}} .
\eqa

Utilizing the above expression, our differential probability becomes

\bqa
\frac{d\mathcal{P}}{D^{3}_{n_{M}}k} &=& G^{2} \frac{2}{\sigma} \iint d\tau' d\tau \;e^{-i\Delta m (\tau' - \tau)} \left| \bra{\prod_{j = 1}^{n_{M}}\mathbf{k}_{j}}  \prod_{\ell = 1}^{n_{M}}\hat{\phi}_{\ell}(x) \ket{0} \right|^{2} \nonumber \\
&=& G^{2}\frac{2}{\sigma}\frac{1}{(2 \pi)^{3n_{M}}}\frac{1}{2^{n_{M}}}  \iint d\tau' d\tau \;e^{-i\Delta m (\tau' - \tau)}   \frac{e^{i\sum_{j = 1}^{n_{M}}(\mathbf{k}_{j} \cdot( \mathbf{x'} -\mathbf{x}) -\omega_{k_{j}} (t'-t))}}{ \prod_{j' = 1}^{n_{M}}  \omega_{j'}}.
\eqa

It should be noted that we are integrating over the accelerated particles proper time. As such, the position and time intervals in the above exponential need to be recast along the trajectory and expressed in terms of the proper time of the accelerated frame. Then, recalling the trajectory from the previous section, Eq. (13), we have

\bqa
\frac{d\mathcal{P}}{D^{3}_{n_{M}}k} &=& G^{2}\frac{2}{\sigma}\frac{1}{(2 \pi)^{3n_{M}}}\frac{1}{2^{n_{M}}} \iint d\tau' d\tau \;e^{-i\Delta m (\tau' - \tau)}   \frac{e^{i\sum_{j = 1}^{n_{M}}(\mathbf{k}_{j} \cdot( \mathbf{x'} -\mathbf{x}) -\omega_{k_{j}} (t'-t))}}{ \prod_{j' = 1}^{n_{M}}  \omega_{j'}} \nonumber \\
&=& G^{2}\frac{2}{\sigma}\frac{1}{(2 \pi)^{3n_{M}}}\frac{1}{2^{n_{M}}}  \iint d\tau' d\tau \;e^{-i\Delta m (\tau' - \tau)}   \frac{ e^{\frac{i}{a}\sum_{j = 1}^{n_{M}}(k_{z_{j}}[\cosh{(a \tau')} - \cosh{(a\tau)}] -\omega_{k_{j}} [\sinh{(a \tau')} - \sinh{(a\tau)}])}}{ \prod_{j' = 1}^{n_{M}}  \omega_{j'}}. 
\eqa

Again, utilizing the change of variables, $u = (\tau' - \tau)/\rho$ and $s = (\tau + \tau')/\sigma$, we recall

\bqa
\cosh{(a\tau')} - \cosh{(a\tau)} &=& 2\sinh{\lb \frac{a\rho u}{2}\rb}\sinh{\lb\frac{a \sigma s}{2} \rb} \nonumber \\
\sinh{(a\tau')} - \sinh{(a\tau)} &=& 2\sinh{\lb\frac{a\rho u}{2}\rb}\cosh{\lb\frac{a \sigma s}{2} \rb}.
\eqa

In changing variables we will again pick up the factor of $\frac{\rho\sigma}{2}$ due to the Jacobian. Using these proper time parametrizations the differential probability becomes

\bqa
\frac{d\mathcal{P}}{D^{3}_{n_{M}}k} &=& G^{2}\frac{2}{\sigma}\frac{1}{(2 \pi)^{3n_{M}}}\frac{1}{2^{n_{M}}}  \iint d\tau' d\tau \;e^{-i\Delta m (\tau' - \tau)}   \frac{ e^{\frac{i}{a}\sum_{j = 1}^{n_{M}}(k_{z_{j}}[\cosh{(a \tau')} - \cosh{(a\tau)}] -\omega_{k_{j}} [\sinh{(a \tau')} - \sinh{(a\tau)}])}}{ \prod_{j' = 1}^{n_{M}}  \omega_{j'}} \nonumber \\
&=& \frac{G^{2}}{(2 \pi)^{3n_{M}}}\frac{\rho}{2^{n_{M}}}  \iint ds du \;e^{-i\Delta m \rho u}   \frac{ e^{\frac{2i}{a}\sum_{j = 1}^{n_{M}}[k_{z_{j}}\sinh{(\frac{a \sigma s}{2})} -\omega_{k_{j}} \cosh{(\frac{a \sigma s}{2})}]\sinh{(\frac{a\rho u}{2})}}}{ \prod_{j' = 1}^{n_{M}}  \omega_{j'}}.
\eqa

Noting that our acceleration is along the $z$ axis only, we can examine the 4-velocity of the accelerated particle using the new affine proper time parametrization $\tilde{s} = \frac{\sigma s}{2}$. Hence

\bqa
u^{\mu}(\tilde{s}) &=& \frac{dx^{\mu}}{d\tilde{s}} \nonumber \\
&=& (\cosh{(a \tilde{s})}, 0, 0, \sinh{(a \tilde{s})}).
\eqa

We can then read off the relativistic factors associated with this motion, $\gamma = \cosh{(a \tilde{s})}$ and $\beta \gamma = \sinh{(a \tilde{s})}$. Then, restricting our analysis to the 2-D subspace along the hyperbolic trajectory, we find that given a 2-momentum $k^{\mu}$ we can boost to the frame instantaneously at rest with the accelerated motion to find

\bqa
\tilde{k}^{\nu} &=& \Lambda^{\nu}_{\mu}k^{\mu} \nonumber \\
&=& \begin{pmatrix}
\gamma & -\beta \gamma \nonumber \\
-\beta \gamma & \gamma
\end{pmatrix}
\begin{pmatrix}
\omega \\
k_{z}
\end{pmatrix} \nonumber \\
&=& \begin{pmatrix}
\cosh{(a \tilde{s})} & -\sinh{(a \tilde{s})} \nonumber \\
-\sinh{(a \tilde{s})} & \cosh{(a \tilde{s})}
\end{pmatrix}
\begin{pmatrix}
\omega \\
k_{z}
\end{pmatrix} \\
\begin{pmatrix}
\tilde{\omega} \\
\tilde{k}_{z}
\end{pmatrix}
&=& \begin{pmatrix}
\omega \cosh{(a \tilde{s})} -k_{z}\sinh{(a \tilde{s})} \\
-\omega \sinh{(a \tilde{s})} + k_{z}\cosh{(a \tilde{s})}
\end{pmatrix}.
\eqa

Upon inspection of the exponential in Eq. (36), we see the argument in the sum is merely the frequency of the emitted particles as seen in the boosted frame instantaneously at rest with accelerated field, i.e. $\tilde{\omega}$. As such we may rewrite the exponential in terms of the boosted frequencies yielding

\bqa
\frac{d\mathcal{P}}{D^{3}_{n_{M}}k} &=& \frac{G^{2}}{(2 \pi)^{3n_{M}}}\frac{\rho}{2^{n_{M}}}  \iint ds du \;e^{-i\Delta m \rho u}   \frac{ e^{\frac{2i}{a}\sum_{j = 1}^{n_{M}}[k_{z_{j}}\sinh{(\frac{a \sigma s}{2})} -\omega_{k_{j}} \cosh{(\frac{a \sigma s}{2})}]\sinh{(\frac{a\rho u}{2})}}}{ \prod_{j' = 1}^{n_{M}}  \omega_{j'}} \nonumber \\
&=& \frac{ G^{2}}{(2 \pi)^{3n_{M}}}\frac{\rho}{2^{n_{M}}}  \iint du ds \;e^{-i\Delta m \rho u}   \frac{ e^{-\frac{2i}{a}[\sum_{j = 1}^{n_{M}}\tilde{\omega}_{k_{j}}]\sinh{(\frac{a\rho u}{2})}}}{ \prod_{j' = 1}^{n_{M}}  \omega_{j'}}.
\eqa

Note the integrand of our differential probability is now independent of the proper time parameter $s$. Therefore we can now divide out the total proper time interval $\int ds = \Delta s$ to obtain the transition probability per unit proper time, $\Gamma_{n_{M}}(\Delta m, a) = \mathcal{P}/\Delta s$. Furthermore, since we have the proper quantity $\tilde{\omega}$ in the exponent we will need to change the remaining momentum variables to the boosted frame as well. Upon inversion of the Lorentz transformations in Eq. (38) we obtain $k_{z} = \tilde{\omega} \sinh{(a\tilde{s})} +\tilde{k}_{z}\cosh{(a\tilde{s})}$ and $\omega = \tilde{\omega} \cosh{(a\tilde{s})}+ \tilde{k}_{z}\sinh{(a\tilde{s})}$. Recalling first that $\tilde{\mathbf{k}}_{\perp} = \mathbf{k}_{\perp}$, we then examine the quantity $dk_{z}/\omega$. Hence

\bqa
\frac{dk_{z}}{\omega} &=& \frac{dk_{z}}{d\tilde{k}_{z}}\frac{d\tilde{k}_{z}}{\omega} \nonumber \\
&=& \frac{d}{d\tilde{k}_{z}}[\tilde{\omega} \sinh{(a\tilde{s})} + \tilde{k}_{z}\cosh{(a\tilde{s})}]\frac{d\tilde{k}_{z}}{\omega} \nonumber \\
&=& [\frac{\tilde{k}_{z}}{\tilde{\omega}} \sinh{(a \tilde{s})} + \cosh{(a \tilde{s})}]\frac{d\tilde{k}_{z}}{\omega} \nonumber \\
&=& \frac{\tilde{k}_{z}\sinh{(a \tilde{s})} + \tilde{\omega}\cosh{(a \tilde{s})}}{\tilde{\omega}} \frac{d\tilde{k}_{z}}{\omega} \nonumber \\
&=&\frac{d\tilde{k}_{z}}{\tilde{\omega}}.
\eqa

The recasting of our transition rate in terms of proper frame variables, accompanied by the rescaling of our proper time via $u \rightarrow \rho u$, yields the following more convenient expression:

\bqa
\Gamma_{n_{M}}(\Delta m, a) &=& \frac{\mathcal{P}}{\Delta s} \nonumber \\
&=& \frac{ G^{2}}{(2 \pi)^{3n_{M}}}\frac{1}{2^{n_{M}}}  \iint du D^{3}_{n_{M}}k  \;e^{-i\Delta m u}   \frac{ e^{-\frac{2i}{a}[\sum_{j = 1}^{n_{M}}\tilde{\omega}_{k_{j}}]\sinh{(\frac{a u}{2})}}}{ \prod_{j' = 1}^{n_{M}}  \omega_{j'}} \nonumber \\
&=& \frac{ G^{2}}{(2 \pi)^{3n_{M}}}\frac{1}{2^{n_{M}}}  \iint du \prod_{\ell = 1}^{n_{M}} d^{3}k_{\ell}  \;e^{-i\Delta m u}   \frac{e^{-\frac{2i}{a}[\sum_{j = 1}^{n_{M}}\tilde{\omega}_{k_{j}}]\sinh{(\frac{a u}{2})}}}{ \prod_{j' = 1}^{n_{M}}  \omega_{j'}} \nonumber \\
&=& \frac{ G^{2}}{(2 \pi)^{3n_{M}}}\frac{1}{2^{n_{M}}}  \iint du \prod_{\ell = 1}^{n_{M}} d^{3}\tilde{k}_{\ell}  \;e^{-i\Delta m u}   \frac{e^{-\frac{2i}{a}[\sum_{j = 1}^{n_{M}}\tilde{\omega}_{k_{j}}]\sinh{(\frac{a u}{2})}}}{ \prod_{j' = 1}^{n_{M}}  \tilde{\omega}_{j'}}.
\eqa

The isotropy of the momentum of the emitted particles in the proper frame is apparent from the above expression. To further facilitate the calculation, we exploit this spherical symmetry by moving our momentum integrations into spherical coordinates. Thus

\bqa
\Gamma_{n_{M}}(\Delta m, a) &=& \frac{ G^{2}}{(2 \pi)^{3n_{M}}}\frac{1}{2^{n_{M}}}  \iint du \prod_{\ell = 1}^{n_{M}} d^{3}\tilde{k}_{\ell}  \;e^{-i\Delta m u}   \frac{ e^{-\frac{2i}{a}[\sum_{j = 1}^{n_{M}}\tilde{\omega}_{k_{j}}]\sinh{(\frac{a u}{2})}}}{ \prod_{j' = 1}^{n_{M}}  \tilde{\omega}_{j'}} \nonumber \\
&=& \frac{ G^{2}}{(2 \pi)^{3n_{M}}}\frac{1}{2^{n_{M}}}  \iint du \prod_{\ell = 1}^{n_{M}}\tilde{k}_{\ell}^{2}\sin{(\tilde{\theta}_{\ell})} d\tilde{k}_{\ell}d\tilde{\theta}_{\ell}d\tilde{\phi}_{\ell}  \;e^{-i\Delta m u}   \frac{ e^{-\frac{2i}{a}[\sum_{j = 1}^{n_{M}}\tilde{\omega}_{k_{j}}]\sinh{(\frac{a u}{2})}}}{ \prod_{j' = 1}^{n_{M}}  \tilde{\omega}_{j'}} \nonumber \\
&=& \frac{ (4 \pi)^{n_{M}}G^{2}}{(2 \pi)^{3n_{M}}}\frac{1}{2^{n_{M}}}  \iint du \prod_{\ell = 1}^{n_{M}}\tilde{k}_{\ell}^{2}d\tilde{k}_{\ell} \;e^{-i\Delta m u}   \frac{ e^{-\frac{2i}{a}[\sum_{j = 1}^{n_{M}}\tilde{\omega}_{k_{j}}]\sinh{(\frac{a u}{2})}}}{ \prod_{j' = 1}^{n_{M}}  \tilde{\omega}_{j'}}.
\eqa

Then, for the final state massless fields $\phi_{i}$, we have $\tilde{\omega}_{i} = \tilde{k}_{i}$ and we may further simplify the above integrations to

\bqa
\Gamma_{n_{M}}(\Delta m, a) &=& \frac{ (4 \pi)^{n_{M}}G^{2}}{(2 \pi)^{3n_{M}}}\frac{1}{2^{n_{M}}}  \iint du \prod_{\ell = 1}^{n_{M}}\tilde{k}_{\ell}^{2}d\tilde{k}_{\ell} \;e^{-i\Delta m u}   \frac{ e^{-\frac{2i}{a}[\sum_{j = 1}^{n_{M}}\tilde{\omega}_{k_{j}}]\sinh{(\frac{a u}{2})}}}{ \prod_{j' = 1}^{n_{M}}  \tilde{\omega}_{j'}} \nonumber \\
&=& G^{2}\frac{ 1}{(2 \pi)^{2n_{M}}}  \iint du \prod_{\ell = 1}^{n_{M}}\tilde{k}_{\ell}^{2}d\tilde{k}_{\ell} \;e^{-i\Delta m u}   \frac{ e^{-\frac{2i}{a}[\sum_{j = 1}^{n_{M}}\tilde{k}_{j}]\sinh{(\frac{a u}{2})}}}{ \prod_{j' = 1}^{n_{M}}  \tilde{k}_{j'}} \nonumber \\
&=& G^{2}\frac{ 1}{(2 \pi)^{2n_{M}}}  \iint du \prod_{\ell = 1}^{n_{M}}\tilde{k}_{\ell}d\tilde{k}_{\ell} \;e^{-i\Delta m u}  e^{-\frac{2i}{a}[\sum_{j = 1}^{n_{M}}\tilde{k}_{j}]\sinh{(\frac{a u}{2})}} \nonumber \\
&=& G^{2}\frac{ 1}{(2 \pi)^{2n_{M}}}  \int du \;e^{-i\Delta m u} \lbk \int d\tilde{k} \;  \tilde{k} e^{-\frac{2i}{a}\tilde{k}\sinh{(\frac{a u}{2})}} \rbk^{n_{M}}.
\eqa

The integral over $\tilde{k}$ will require the use of a regulator to the ensure convergence of the integral. In order to damp the oscillation at infinity, we let $\sinh{(\frac{a u}{2})} \rightarrow \sinh{(\frac{a u}{2})}-i\epsilon \approx \sinh{(\frac{a u}{2}-i\epsilon)}$ with $\epsilon > 0$. As such, the momentum integration yields

\bqa
\int d\tilde{k} \;  \tilde{k} e^{-\frac{2i}{a}\tilde{k}\sinh{(\frac{a u}{2})}} &=& \int_{0}^{\infty} d\tilde{k} \;  \tilde{k} e^{-\frac{2i}{a}\tilde{k}(\sinh{(\frac{a u}{2})} -i\epsilon)} \nonumber \\
&=& \lbk e^{-\frac{2i}{a}\tilde{k}(\sinh{(\frac{a u}{2})} -i\epsilon)}\frac{(1+\frac{2}{a}\tilde{k}(\sinh{(\frac{a u}{2})} -i\epsilon))}{(\frac{2}{a}(\sinh{(\frac{a u}{2})} -i\epsilon))^{2}} \rbk^{\infty}_{0} \nonumber \\
&=&  -\frac{a^{2}}{4}\frac{1}{\sinh^{2}{(\frac{a u}{2} -i\epsilon)}}. 
\eqa

It should be noted that, up to a multiplicative constant, we have reproduced the Wightman function for a massless scalar field in Rindler space, Eq. (16). We arrived at this expression by inserting the hyperbolic trajectory into the mode functions prior to evaluating the two point function rather than evaluating the two point function first and then inserting the trajectory as we did in the previous section. The fact that we obtained the same result serves as a self-consistency check. This method also served to shed light on the physics of the emission process in the proper frame. For a more comprehensive analysis of the physics of the proper frame we refer the reader to Ref. [6]. Our acceleration induced transition rate, Eq. (43), then takes the form 

\bqa
\Gamma_{n_{M}}(\Delta m, a) &=& G^{2}\frac{ 1}{(2 \pi)^{2n_{M}}}  \int du \;e^{-i\Delta m u} \lbk \int d\tilde{k} \;  \tilde{k} e^{-\frac{2i}{a}\tilde{k}\sinh{(\frac{a u}{2})}} \rbk^{n_{M}} \nonumber \\
&=& G^{2}\frac{ 1}{(2 \pi)^{2n_{M}}}  \int du \;e^{-i\Delta m u} \lbk -\frac{a^{2}}{4}\frac{1}{\sinh^{2}{(\frac{a u}{2} -i\epsilon)}}\rbk^{n_{M}} \nonumber \\
&=& G^{2}\lb \frac{ ia}{4 \pi}\rb^{2n_{M}}  \int du \;  \frac{e^{-i\Delta m u}}{[\sinh{(\frac{a u}{2} -i\epsilon)}  ]^{2n_{M}}}.
\eqa

A similar integral, Eq. (18), was encountered in the previous section. By making the replacement in the integrand $m\rightarrow -\Delta m$ we can quote the result by inspection. Hence,

\bqe
\Gamma_{n_{M}}(\Delta m, a) = G^{2}\lb \frac{ia}{2 \pi} \rb^{2n_{M}} \frac{1}{a} \frac{2\pi i}{(2n_{M}-1)!} \frac{\Gamma(-i\Delta m/a+n_{M})}{\Gamma(-i \Delta m/a +1-n_{M})} \frac{1}{1- e^{2\pi  \Delta m/a} }.
\eqe 

To recast the above gamma functions into the same form as the previous section we recall the identity $\Gamma(z)\Gamma(1-z) =  \frac{\pi}{\sin{(\pi z)}}$ to find

\bqa
 \frac{\Gamma(-i\Delta m/a+n_{M})}{\Gamma(-i \Delta m/a +1-n_{M})} = -\frac{\Gamma(i\Delta m/a+n_{M})}{\Gamma(i \Delta m/a +1-n_{M})}.
\eqa

Thus, our total rate, Eq. (46), of our two-level system to undergo an acceleration induced transition and simultaneously emit $n_{M}$ massless scalar fields is given by

\bqa
\Gamma_{n_{M}}(\Delta m, a) = G^{2}\lb \frac{ia}{2 \pi} \rb^{2n_{M}} \frac{1}{a} \frac{2\pi i}{(2n_{M}-1)!} \frac{\Gamma(i\Delta m/a+n_{M})}{\Gamma(i \Delta m/a +1-n_{M})}\frac{1}{e^{2\pi  \Delta m/a} -1 }.
\eqa

As expected, we have reproduced the same expression as in the previous section provided we made the appropriate identifications for $m$. The use of an Unruh-DeWitt detector has provided us with a relatively simple procedure for including a massive particle in the final state but at the expense of keeping it confined to Rindler space. This is due to one of the final state particles being locked in the detector. Again, normalizing the transition rate via $\tilde{\Gamma} = \Gamma/\Gamma_{0}$ with $\Gamma_{0} = G^{2}$, we write out the first few normalized decay rates $\tilde{\Gamma}_{n_{M}}$. Hence,

\bqa
\tilde{\Gamma}_{1}(\Delta m,a) &=& \frac{\Delta m }{2 \pi} \frac{1}{e^{2\pi\Delta m/a}-1} \non \\
\tilde{\Gamma}_{2}(\Delta m,a) &=& \frac{(\Delta m)^{3}}{48 \pi^{3}} \frac{1+\lb \frac{a}{\Delta m}  \rb^{2}}{e^{2\pi\frac{\Delta m}{a}}-1} \non \\
\tilde{\Gamma}_{3}(\Delta m,a) &=& \frac{(\Delta m)^{5}}{3840 \pi^{5}} \frac{1+ 5 \lb \frac{a}{\Delta m}  \rb^{2} + 4 \lb \frac{a}{\Delta m}  \rb^{4}}{e^{2\pi\Delta m/a}-1} \non \\
\tilde{\Gamma}_{4}(\Delta m,a) &=& \frac{(\Delta m)^{7}}{645120 \pi^{7}} \frac{1+ 14 \lb \frac{a}{\Delta m}  \rb^{2} + 49 \lb \frac{a}{\Delta m}  \rb^{4} + 36 \lb \frac{a}{\Delta m}  \rb^{6}}{e^{2\pi\Delta m/a}-1} \non \\
\tilde{\Gamma}_{5}(\Delta m,a) &=& \frac{(\Delta m)^{9}}{185794560 \pi^{9}} \frac{1+ 30 \lb \frac{a}{\Delta m}  \rb^{2} + 273 \lb \frac{a}{\Delta m}  \rb^{4} + 820 \lb \frac{a}{\Delta m}  \rb^{6} + 576 \lb \frac{a}{\Delta m}  \rb^{8} }{e^{2\pi\Delta m/a}-1}.
\eqa

Comparing with the previous rates from Eq. (22), the use of an Unruh-DeWitt detector to model particle decays produces a similar form for the decay rate but with a more general mass transition. This is due to the fact that the particle that is coupled into the two-level system with the initial accelerated particle remains in Rindler space. In the previous section all final state particles were emitted into Minkowski space and it was the Wightman functions of these particles which contributed to the polynomial. Therefore one must take care when analyzing a system to ensure that the final state particles, i.e. fields, are expressed in terms of the mode functions of the appropriate spacetime.

By using the Unruh-DeWitt detector we can analyze not only acceleration induced decays but also excitations. By letting $\Delta m = -1$ we can reproduce the results of the previous section. Rather we set $\Delta m = 1$ to analyze an initially accelerated particle that excites into a more massive state. We can now look at normalized $\tilde{\Gamma}_{n_{M}} $ detector excitation rates with the simultaneous emission of $n_{M}$ massless particles into Minkowski space. We focus this analysis for $a \geq \Delta m$ since the relevant plots rapidly diverge at low acceleration to reflect the infinite lifetimes for stable particles in inertial frames (see Figs. 3 and 4).

\begin{figure}[H]
\centering  
\includegraphics[,scale=.6]{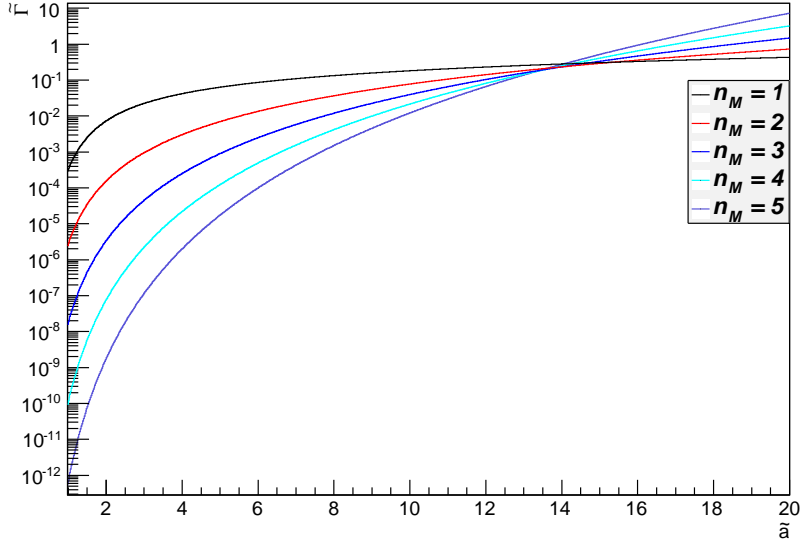}
\caption{The normalized excitation rates, Eq. (49), with $\tilde{a} = a/\Delta m$ and $\Delta m = 1$.}
\end{figure}

\begin{figure}[H]
\centering  
\includegraphics[,scale=.6]{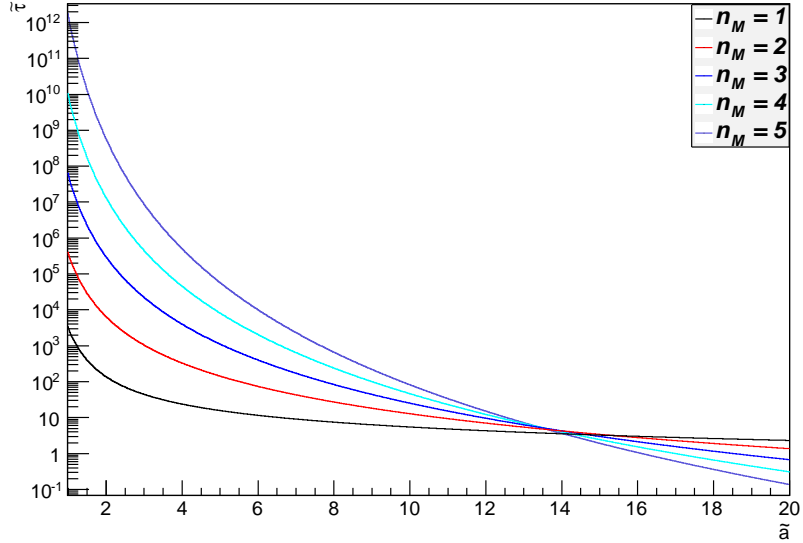}
\caption{The normalized excitation lifetimes $\tilde{\tau} = 1/\tilde{\Gamma}$ with $\tilde{a} = a/\Delta m$ and $\Delta m = 1$.}
\end{figure}

We have found in this section that the use of an Unruh-DeWitt detector allows for a more general mass transition when examining the effect of acceleration on unstable particles. This is due to the coupling of one of the final state products into the accelerated detector which effectively keeps this particle in Rindler space. This situation arises, for example, when the acceleration mechanism is an electric field and an initial charged particle undergoes a transition into another charged particle with the simultaneous emission of two neutral particles. The final state charged particle remains in Rindler space on account of the acceleration due to the electric field while the neutral particles are unaffected by the electric field and are thus effectively in Minkowski space. A muon accelerated by an electric field and decaying into an electron and two neutrinos is an example of this type of process. This and the reverse process of electron excitation will be analyzed in a later section. In the next section we generalize the accelerated field transition process to arbitrary final state multiplicities in both Rindler and Minkowski spacetimes.

\section{Generalized $N$ Particle Scalar Multiplicities}

In the previous sections we evaluated the acceleration induced transition rate using two different methods. We now demonstrate the equivalence between the two methods and also show how to correctly interpret and make use of the overall formalism considering an initial particle in Rindler spacetime and allowing it to decay into $n_{R}$ particles in Rindler space and $n_{M}$ particles into Minkowski space. Schematically we are examining the process

\bqa
\Psi_{i} \rightarrow_{a} \Psi_{1} \Psi_{2} \cdots \Psi_{n_{R}} \phi_{1} \phi_{2} \cdots \phi_{n_{M}}
\eqa

We denote the initial accelerated massive field by $\Psi_{i}$, the final state Rindler particles of arbitrary mass by $\Psi_{j}$, and the massless final state Minkowski particles by $\phi_{k}$. In order to analyze this process we consider the following more general interaction action:

\bqe
\hat{S}_{I} = \int d^{4}x\sqrt{-g} \sqrt{\frac{2}{\sigma \kappa}}G \hat{\Psi}_{i} \prod_{r = 1}^{n_{R}}\hat{\Psi}_{r} \prod_{m = 1}^{n_{M}}\hat{\phi}_{m}.
\eqe

As before, the coupling constant $G$ will be determined by the inertial limit of the specific interaction in question and the additional factor $\sqrt{\frac{2}{\sigma \kappa}}$ is defined for later convenience. The probability amplitude for the acceleration induced transition of our massive initial state into $n$ total particles is given by

\bqe
\mathcal{A} = \bra{\prod_{\ell = 1}^{n_{M}}\mathbf{k}_{\ell}}\otimes \bra{\prod_{j = 1}^{n_{R}}\Psi_{j}} \hat{S}_{I}\ket{\Psi_{i}} \otimes \ket{0}.
\eqe

The Rindler states $\ket{\Psi_{j}}$ are labeled by the index $j$ while the Minkowski states $\ket{\mathbf{k}_{\ell}}$ are labeled by their momenta. We again use the same notation for our Fock states and accommodate any complications due to the statistics or degeneracies of the final state products by rescaling our effective coupling. With the same notation $\prod_{n = 1}^{n_{M}} d^{3}k_{n_{M}} = D^{3}_{n_{M}}k$ the differential probability for the accelerated field to decay and emit $n_{R}$ particles into Rindler space and $n_{M}$ particles into Minkowski space is given by

\bqa
\frac{d \mathcal{P}}{D^{3}_{n_{M}}k} &=&  |\mathcal{A}|^{2} \nonumber \\
&=&  G^{2}\frac{2}{\sigma \kappa}\left| \int d^{4}x\sqrt{-g}\bra{\prod_{\ell = 1}^{n_{M}}\mathbf{k}_{\ell}}\otimes \bra{\prod_{j = 1}^{n_{R}}\Psi_{j}}   \hat{\Psi}_{i}(x) \prod_{r = 1}^{n_{R}}\hat{\Psi}_{r}(x) \prod_{m = 1}^{n_{M}}\hat{\phi}_{m}(x) \ket{\Psi_{i}} \otimes \ket{0} \right|^{2} \nonumber \\
&=& G^{2}\frac{2}{\sigma \kappa}\left| \int d^{4}x\sqrt{-g} \bra{\prod_{j = 1}^{n_{R}}\Psi_{j}}   \hat{\Psi}_{i}(x) \prod_{r = 1}^{n_{R}}\hat{\Psi}_{r}(x) \ket{\Psi_{i}} \bra{\prod_{\ell = 1}^{n_{M}}\mathbf{k}_{\ell}} \prod_{m = 1}^{n_{M}}\hat{\phi}_{m}(x)   \ket{0} \right|^{2} \non \\
&=& G^{2} \frac{2}{\sigma \kappa} \iint d^{4}x d^{4}x' \sqrt{-g}\sqrt{-g'} \left|\bra{\prod_{j = 1}^{n_{R}}\Psi_{j}}   \hat{\Psi}_{i}(x) \prod_{r = 1}^{n_{R}}\hat{\Psi}_{r}(x) \ket{\Psi_{i}}\right|^{2} \left| \bra{\prod_{\ell = 1}^{n_{M}}\mathbf{k}_{\ell}} \prod_{m = 1}^{n_{M}}\hat{\phi}_{m}(x)   \ket{0} \right|^{2}.
\eqa  

We can now factor out the $n_{M}$ complete set of momentum eigenstates. The result will give the product of Wightman functions of the massless Minkowski fields,

\bqa
\mathcal{P} &=&  G^{2} \frac{2}{\sigma \kappa} \iint d^{4}x d^{4}x' \sqrt{-g}\sqrt{-g'} \left|\bra{\prod_{j = 1}^{n_{R}}\Psi_{j}}   \hat{\Psi}_{i}(x) \prod_{r = 1}^{n_{R}}\hat{\Psi}_{r}(x) \ket{\Psi_{i}}\right|^{2} \prod_{n = 1}^{n_{M}} \int d^{3}k_{n_{M}} \left| \bra{\prod_{\ell = 1}^{n_{M}}\mathbf{k}_{\ell}} \prod_{m = 1}^{n_{M}}\hat{\phi}_{m}(x)   \ket{0} \right|^{2} \non \\
&=&  G^{2} \frac{2}{\sigma \kappa} \iint d^{4}x d^{4}x' \sqrt{-g}\sqrt{-g'} \left|\bra{\prod_{j = 1}^{n_{R}}\Psi_{j}}   \hat{\Psi}_{i}(x) \prod_{r = 1}^{n_{R}}\hat{\Psi}_{r}(x) \ket{\Psi_{i}}\right|^{2}  \prod_{m = 1}^{n_{M}} \bra{0}\hat{\phi}_{m}(x') \hat{\phi}_{m}(x)   \ket{0}\non \\
&=& G^{2}  \frac{2}{\sigma \kappa} \iint d^{4}x d^{4}x' \sqrt{-g}\sqrt{-g'} \left|\bra{\prod_{j = 1}^{n_{R}}\Psi_{j}}   \hat{\Psi}_{i}(x) \prod_{r = 1}^{n_{R}}\hat{\Psi}_{r}(x) \ket{\Psi_{i}}\right|^{2}  [G^{\pm}(x',x)]^{n_{M}}.
\eqa

We now examine the remaining Rindler space inner products. As before, we have seen that each field operator serves to extract the appropriate mode function of each Rindler particle. The Rindler coordinate proper time of the initial field will again serve as our time coordinate. As such we can examine the above inner products. Hence

\bqa
\bra{\prod_{j = 1}^{n_{R}}\Psi_{j}}   \hat{\Psi}_{i}(x) \prod_{r = 1}^{n_{R}}\hat{\Psi}_{r}(x) \ket{\Psi_{i}} &=& f_{\Psi_{i}}[x(\tau)]e^{-im_{i}\tau} \prod_{r = 1}^{n_{R}} f_{\Psi_{r}}^{\ast}[x(\tau)]e^{i\omega_{r}\tau} \non \\
&=& \left[f_{\Psi_{i}}[x(\tau)] \prod_{r = 1}^{n_{R}} f_{\Psi_{r}}^{\ast}[x(\tau)]\right] e^{i\Delta E_{R}\tau}.
\eqa 

The Rindler mode frequencies $\omega_{r}$ correspond to the energies of final state Rindler particles which may not necessarily be the appropriate rest masses. Also we have defined $\Delta E_{R} = \sum \omega_{r} - m_{i}$ to be the total energy difference between the final and initial Rindler space field configuration. Our total transition probability, Eq. (54), then becomes

\bqa
\mathcal{P} &=& G^{2} \frac{2}{\sigma \kappa} \iint d^{4}x d^{4}x' \sqrt{-g}\sqrt{-g'} \left|\bra{\prod_{j = 1}^{n_{R}}\Psi_{j}}   \hat{\Psi}_{i}(x) \prod_{r = 1}^{n_{R}}\hat{\Psi}_{r}(x) \ket{\Psi_{i}}\right|^{2}  [G^{\pm}(x',x)]^{n_{M}} \non \\
&=& G^{2} \frac{2}{\sigma \kappa} \iint d^{4}x d^{4}x' \sqrt{-g}\sqrt{-g'} \left|\left[f_{\Psi_{i}}[x(\tau)] \prod_{r = 1}^{n_{R}} f_{\Psi_{r}}^{\ast}[x(\tau)]\right] e^{i\Delta E_{R}\tau}\right|^{2}  [G^{\pm}(x',x)]^{n_{M}} \non \\
&=& G^{2} \frac{2}{\sigma \kappa} \iint d^{4}x d^{4}x' \sqrt{-g}\sqrt{-g'} \left|f_{\Psi_{i}}[x(\tau)] \prod_{r = 1}^{n_{R}} f_{\Psi_{r}}^{\ast}[x(\tau)] \right|^{2} e^{-i\Delta E_{R}(\tau'-\tau)} [G^{\pm}(x',x)]^{n_{M}}.
\eqa

We again define $\kappa$ to be the overall normalization of the product of envelope functions $f_{\Psi}$, i.e.

\bqa
\kappa = \iint d^{3}x d^{3}x' \sqrt{-g}\sqrt{-g'} \left|f_{\Psi_{i}}[x(\tau)] \prod_{r = 1}^{n_{R}} f_{\Psi_{r}}^{\ast}[x(\tau)] \right|^{2}.
\eqa

As such the total probability for our transition becomes

\bqa
\mathcal{P} &=& G^{2} \frac{2}{\sigma } \iint d\tau d\tau'  e^{-i\Delta E_{R}(\tau'-\tau)} [G^{\pm}(x',x)]^{n_{M}}.
\eqa

In carrying out this analysis we see that one can consider having a transition involving an arbitrary number of final state particles in Rindler space to be equivalent to having an Unruh-DeWitt detector with the energy levels being the initial and final state energies of the Rindler space field configuration as seen in the proper frame of the initially accelerated field. Having evaluated this expression before we know the remaining procedures are to formulate the transition rate and evaluate the Fourier transform of the product of the Minkowski final state Wightman functions evaluated along the accelerated trajectory of the initial Rindler particle state. We now quote the final form of the transition probability. Thus

\bqa
\Gamma_{n_{M}}(\Delta E_{R}, a) &=& G^2 \lb \frac{ia}{2 \pi}  \rb^{2n_{M}} \frac{1}{a} \frac{2 \pi i}{(2n_{M}-1)!}\frac{\Gamma(i\Delta E_{R}/a+n_{M})}{\Gamma(i\Delta E_{R}/a+1-n_{M})}\frac{1}{e^{2\pi\Delta E_{R}/a}-1}.
\eqa 

This is the same form of the expression that we have arrived at previously but now we have a clearer understanding of the role each of the Rindler and Minkowski space fields plays in the transition rate. For the sake of completeness, we list the normalized decay rates $\tilde{\Gamma}_{n_{M}}(\Delta E_{R}, a)$, for the first few multiplicities. Hence,

\bqa
\tilde{\Gamma}_{1}(\Delta E_{R}, a) &=& \frac{\Delta E_{R}}{2 \pi} \frac{1}{e^{2\pi\Delta E_{R}/a}-1} \non \\
\tilde{\Gamma}_{2}(\Delta E_{R}, a) &=& \frac{\Delta E_{R}^{3}}{48 \pi^{3}} \frac{1+\lb \frac{a}{\Delta E_{R}}  \rb^{2}}{e^{2\pi\Delta E_{R}/a}-1} \non \\
\tilde{\Gamma}_{3}(\Delta E_{R}, a) &=& \frac{\Delta E_{R}^{5}}{3840 \pi^{5}} \frac{1+ 5 \lb \frac{a}{\Delta E_{R}}  \rb^{2} + 4 \lb \frac{a}{\Delta E_{R}}  \rb^{4}}{e^{2\pi\Delta E_{R}/a}-1} \non \\
\tilde{\Gamma}_{4}(\Delta E_{R}, a) &=& \frac{\Delta E_{R}^{7}}{645120 \pi^{7}} \frac{1+ 14 \lb \frac{a}{\Delta E_{R}}  \rb^{2} + 49 \lb \frac{a}{\Delta E_{R}}  \rb^{4} + 36 \lb \frac{a}{\Delta E_{R}}  \rb^{6}}{e^{2\pi\Delta E_{R}/a}-1} \non \\
\tilde{\Gamma}_{5}(\Delta E_{R}, a) &=& \frac{\Delta E_{R}^{9}}{185794560 \pi^{9}} \frac{1+ 30 \lb \frac{a}{\Delta E_{R}}  \rb^{2} + 273 \lb \frac{a}{\Delta E_{R}}  \rb^{4} + 820 \lb \frac{a}{\Delta E_{R}}  \rb^{6} + 576 \lb \frac{a}{\Delta E_{R}}  \rb^{8} }{e^{2\pi\Delta E_{R}/a}-1}.
\eqa 

The difficulty in measuring these effects is that the acceleration scale currently accessible in laboratory settings is significantly smaller than the energy scale of the transition. If, through some mechanism, we could not only control the acceleration but also the transition energy scale we could bring the effects closer to our experimental reach. A mathematical analysis of the energy spectra of Rindler particles which have decay products in both Rindler and Minkowski spacetime has yet to be carried out but would provide a much clearer insight into the how any Rindler particle energies would be perceived in the proper frame of the accelerated field. With this in mind we plot, in Figs. 5 and 6, the normalized decay rates and lifetimes for a constant acceleration $a = 1$ while varying the energy scale $\Delta E_{R}$. 

\begin{figure}[H]
\centering  
\includegraphics[,scale=.6]{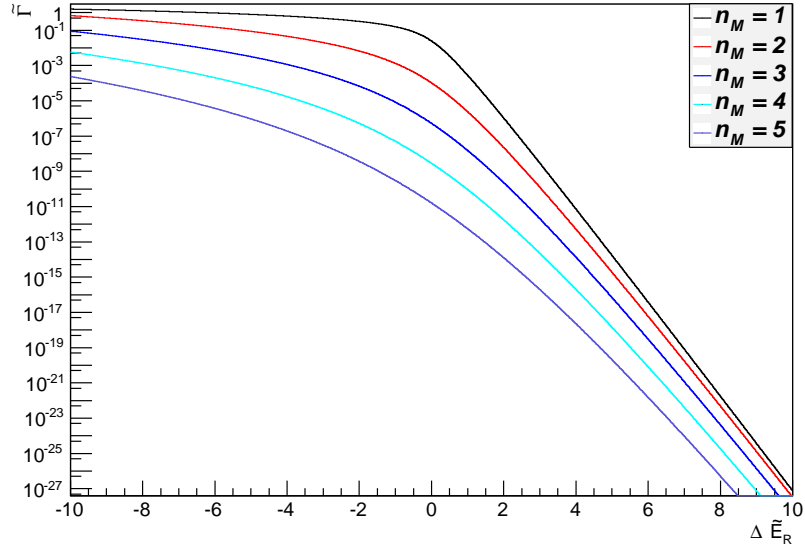}
\caption{The normalized transition rates, Eq. (60), with $\Delta \tilde{E}_{R} = \Delta \tilde{E}_{R}/a$ and $a=1$.}
\end{figure}

\begin{figure}[H]
\centering  
\includegraphics[,scale=.6]{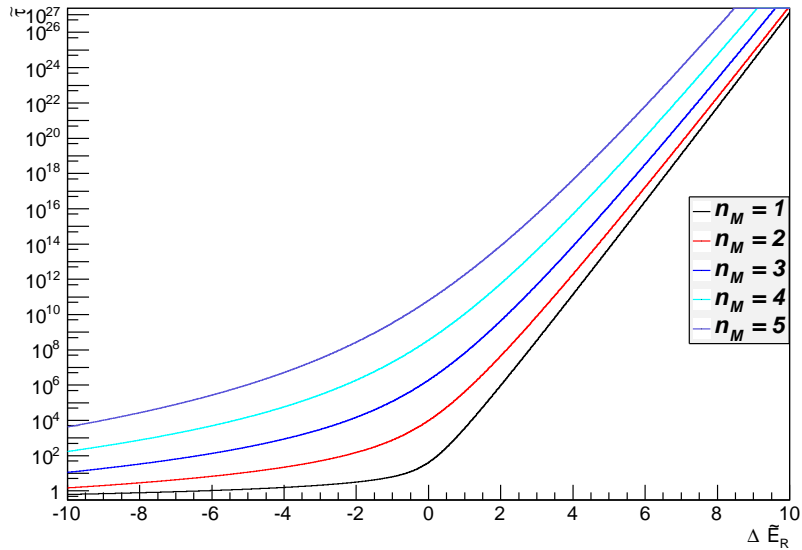}
\caption{The normalized transition lifetimes $\tilde{\tau} = 1/\tilde{\Gamma}$ with $\Delta \tilde{E}_{R} = \Delta \tilde{E}_{R}/a$ and $a=1$.}
\end{figure}

To better understand the role each spacetime field configuration has in the transition rate, we define the polynomial of multiplicity $\mathcal{M}_{n_{M}}(\Delta E_{R}, a)$ as follows:

\bqe
\mathcal{M}_{n_{M}}(\Delta E_{R}, a) = \lb \frac{ia}{2 \pi}  \rb^{2n_{M}} \frac{1}{a} \frac{2 \pi i}{(2n_{M}-1)!}\frac{\Gamma(i\Delta E_{R}/a+n_{M})}{\Gamma(i\Delta E_{R}/a+1-n_{M})}.
\eqe

We then find the general form for the decay rate to be

\bqa
\Gamma_{n_{M}}(\Delta E_{R}, a) &=& G^2\mathcal{M}_{n_{M}}(\Delta E_{R}, a)f(\Delta E_{R}, a).
\eqa

We see that the rate factors into the inertial interaction specific coupling, the polynomial of multiplicity, and the thermal distribution $f(\Delta E_{R}, a)$ associated with the Unruh effect. The final state multiplicity in Minkowski space governs the number of terms in the polynomial while the total change of the energy in Rindler space sets the acceleration scale of the transition rate. The inertial interaction coupling constant sets the overall normalization of the transition rate.

In this section we generalized the analysis of acceleration induced field transitions to that of arbitrary Rindler and Minkowski space particle multiplicities. We determined the roles that each spacetime field configuration plays in the transition rate and examined how the rates evolve with the total energy change of Rindler space at constant acceleration. The next section focuses on the application of the above formalism to that of the electron and muon system.

\section{The Electron and Muon System}

The weak decay of muons into electrons could possibly provide a robust setting to investigate the effects of acceleration on certain aspects of the physics of unstable particles. We will apply the results of the previous section to model both muon decay as well as the reverse process of electron excitation utilizing the scalar field approximation. In addition to the standard decay/excitation rates, we will also compute the branching fractions of the muon decay chains as a function of proper acceleration. To model the muon and electron transitions we will assume the acceleration mechanism is something like an electric field so that both the muon and the electron are effectively in Rindler space, due to their charge, while the neutral neutrinos are emitted into Minkowski space. This setup more closely resembles an actual experimental setting which could, in principle, investigate this phenomena. Schematically we will analyze the following processes:

\bqa
\mu^{\pm} \rightarrow_{a} e^{\pm} + \bar{\nu}_{e} + \nu_{\mu}, \;\;\; e^{\pm} \rightarrow_{a} \mu^{\pm} + \bar{\nu}_{\mu} + \nu_{e}.
\eqa 

The transition rate which describes both of these processes is given by the $n_{M} = 2$ case from Eq. (60),

\bqa
\Gamma^{\mu \leftrightarrow e}(\Delta E_{R},a) = G^2 \frac{(\Delta E_{R})^{3}}{48 \pi^{3}} \frac{1+\lb \frac{a}{\Delta E_{R}}  \rb^{2}}{e^{2\pi\Delta E_{R}/a}-1}.
\eqa 

To determine the coupling constant $G$ we compare the inertial limit of the above accelerated decay rate to that of the known inertial muon decay rate. The known decay rate of inertial muons, to lowest order in perturbation theory [12], is given by 

\bqe
\Gamma_{i}^{\mu} = \frac{G^{2}_{f}m^{5}_{\mu}}{192 \pi^{3}},
\eqe

where $G_{f}$ is the Fermi coupling constant. Note we have disregarded higher order terms which contain powers of $m_{e}/m_{\mu}$. As such, in our analysis of muon decay we may consider the electron, and of course the neutrinos, to be massless. In addition to considering the electron to be massless, we will also assume that the total energy of the electron emitted into Rindler space will be insignificant when compared to the muon mass. The specrta of the final state Minkowski particles has been calculated in Ref. [6] and indicates that each particle will have an energy distribution, as measured in the inertial frame instantaneously at rest with the initial accelerated particle, peaked about the proper acceleration. A computation of the energy spectra with the appropriate particles emitted into Rindler space has yet to be carried out. This  would help more accurately determine the final state electron energy associated with the decay of accelerated muons. Recalling that $\Delta E_{R} = \sum \omega_{R} - m_{i}$, we will then have $\Delta E_{R} = -m_{\mu}$ for the current analysis. By taking the limit $a \rightarrow 0$ of the acceleration induced decay rate, Eq. (64), and equating it with the known inertial decay rate, Eq. (65), we can determine our effective coupling constant. Thus,

\bqa
\lim_{a \rightarrow 0} \Gamma^{\mu \rightarrow e}(\Delta E_{R}, a) &=& \Gamma_{i}^{\mu} \nonumber \\
\lim_{a \rightarrow 0} G^2 \frac{m_{\m}^{3}}{48 \pi^{3}} \frac{1+\lb \frac{a}{m_{\m}}  \rb^{2}}{1-e^{-2\pi m_{\m}/a}}&=& \frac{G^{2}_{f}m^{5}_{\mu}}{192 \pi^{3}} \nonumber \\
\frac{G^2 m_{\mu}^{3}}{48\pi^{3}} &=& \frac{G^{2}_{f}m^{5}_{\mu}}{192 \pi^{3}} \nonumber \\
G &=& \frac{1}{2}m_{\mu}G_{f}.
\eqa

As such, the properly normalized muon decay rate under the influence of acceleration is given by

\bqe
\Gamma^{\mu \rightarrow e}(a) =  \frac{G^{2}_{f}m^{5}_{\mu}}{192 \pi^{3}}\frac{ 1 +\lb \frac{a}{m_{\mu}} \rb^{2} }{1-e^{-2\pi m_{\m}/a}}.
\eqe

Our result differs from that of Mueller [4] by having a lower order polynomial due to our assumption of keeping the final state electron in Rindler space. Had we allowed the electron to be created in Minkowski space we would have recovered the same result as Mueller. Furthermore, the inclusion of fermions in the analysis would also yield a higher order polynomial due to the additional factors of frequency in the standard fermionic normalization [5-7]. This yields higher powers of frequency to be integrated over when summing over the final state momentum of the Minkowski particles. In either case, the resultant expressions are equivalent at low accelerations but also illustrate the fact that at high accelerations one needs to be precise in describing such processes. By recalling that $G_{f} = 1.166 \times 10^{-5} \; \text{GeV}^{-2}$ and $m_{\mu} = 105.7 \; \text{MeV}$, we can evaluate the canonical inertial muon lifetime $\frac{192 \pi^{3}}{G^{2}_{f}m^{5}_{\mu}} = \tau_{\mu} = 2.184 \; \mu \text{s}$ which sets the overall scale of our transition rate. We need to also mention that the energy scale of the interaction is set by the acceleration. In this analysis we are using a scalar approximation of an effective Fermi interaction. The nonrenormalizability of this approximation necessitates the interaction energy to be less than the rest masses of the weak gauge bosons. With masses $m_{W}, m_{Z} \sim 1000 \; \text{GeV}$ we have carried out all our analysis with accelerations from 0 to 20 in muon mass units. With the muon system under consideration we have $\frac{20m_{\m}}{m_{W},m_{Z}} \sim .01 \ll 1$ and therefore our analysis remains valid. Plots of the acceleration-dependent muon decay rate and lifetime are shown below (see Figs. 7 and 8).

\begin{figure}[H]
\centering  
\includegraphics[,scale=.6]{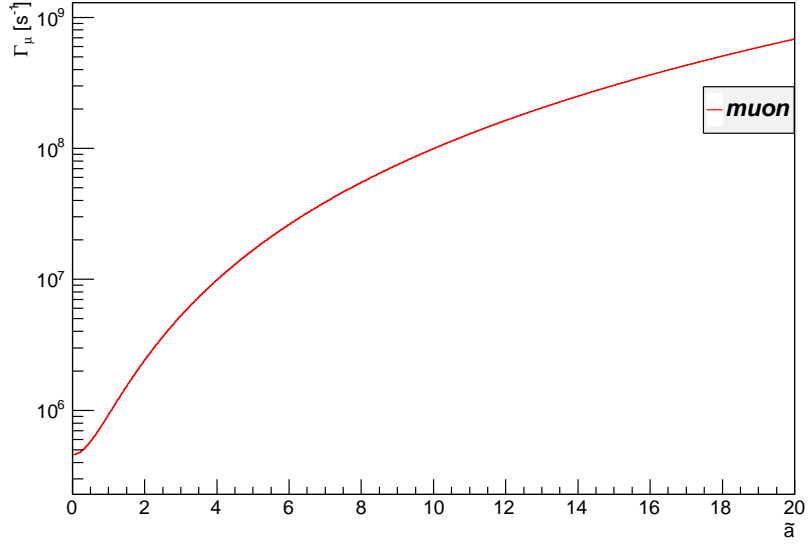}
\caption{The muon decay rate, Eq. (67), as a function of $\tilde{a} = \frac{a}{m_{\mu}}$.}
\end{figure}

\begin{figure}[H]
\centering  
\includegraphics[,scale=.6]{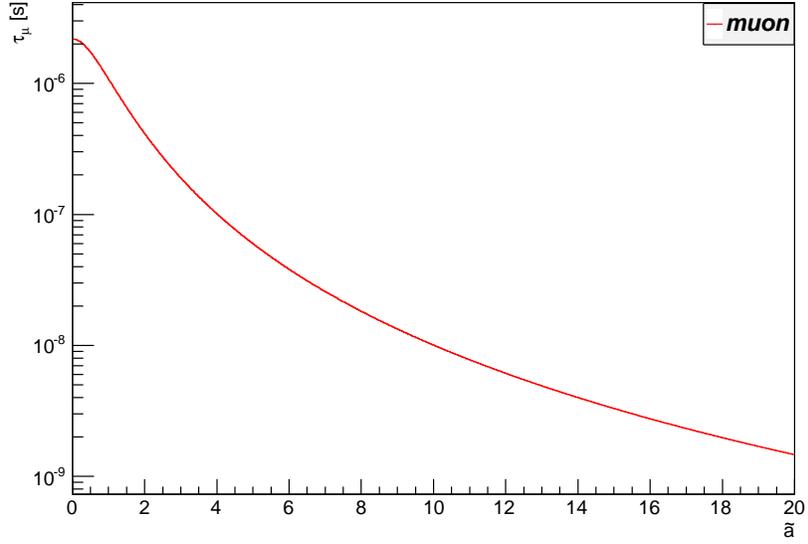}
\caption{The muon lifetime $\tau_{\m} = 1/\Gamma_{\m}$ as a function of $\tilde{a} = \frac{a}{m_{\mu}}$.}
\end{figure}

We now apply the transition rate, Eq. (64), to the case of electron excitation. To do so, we use the detailed balance between the transitions $\Gamma^{e \rightarrow \m} = e^{-2\pi\Delta E_{R}/a} \Gamma^{\m \rightarrow e}$ at thermal equilibrium [10]. This is affected by merely reversing the sign $\Delta E_{R} \rightarrow -\Delta E_{R}$ or rather we take $m_{\m} \rightarrow -m_{\m}$ in Eq. (67). This also enables us to keep the overall coupling constant from the muon decay by using the symmetry between the two thermalized processes. Furthermore, this implies that the Rindler space energy of the created muon comprises mainly the mass with no appreciable momentum. Again we note that a better understanding of the energy spectra of all particles in all spacetimes is necessary to more accurately model these processes. With these considerations we can now estimate the acceleration induced excitation of electrons back into muons to be

\bqe
\Gamma^{e \rightarrow \m}(a) =  \frac{G^{2}_{f}m^{5}_{\mu}}{192 \pi^{3}}\frac{ 1 +\lb \frac{a}{m_{\mu}} \rb^{2} }{e^{2\pi m_{\m}/a}-1}.
\eqe

We can now plot, in Figs. 9 and 10, the excitation rate as well as the lifetime. Note the fact that the decay rate rapidly approaches zero as $a \rightarrow 0$, and thus causes the lifetime to diverge. This reflects the stability of inertial electrons.

\begin{figure}[H]
\centering  
\includegraphics[,scale=.6]{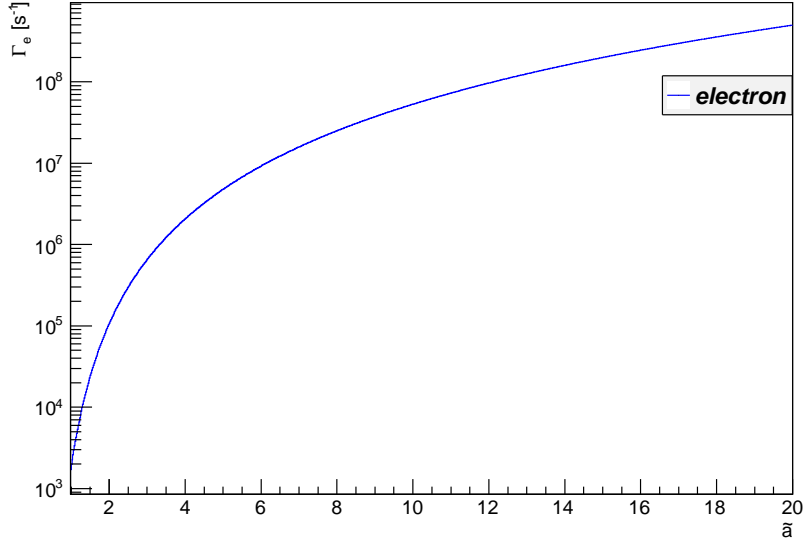}
\caption{The electron excitation rate, Eq. (68), as a function of $\tilde{a} = \frac{a}{m_{\m}}$.}
\end{figure}

\begin{figure}[H]
\centering  
\includegraphics[,scale=.6]{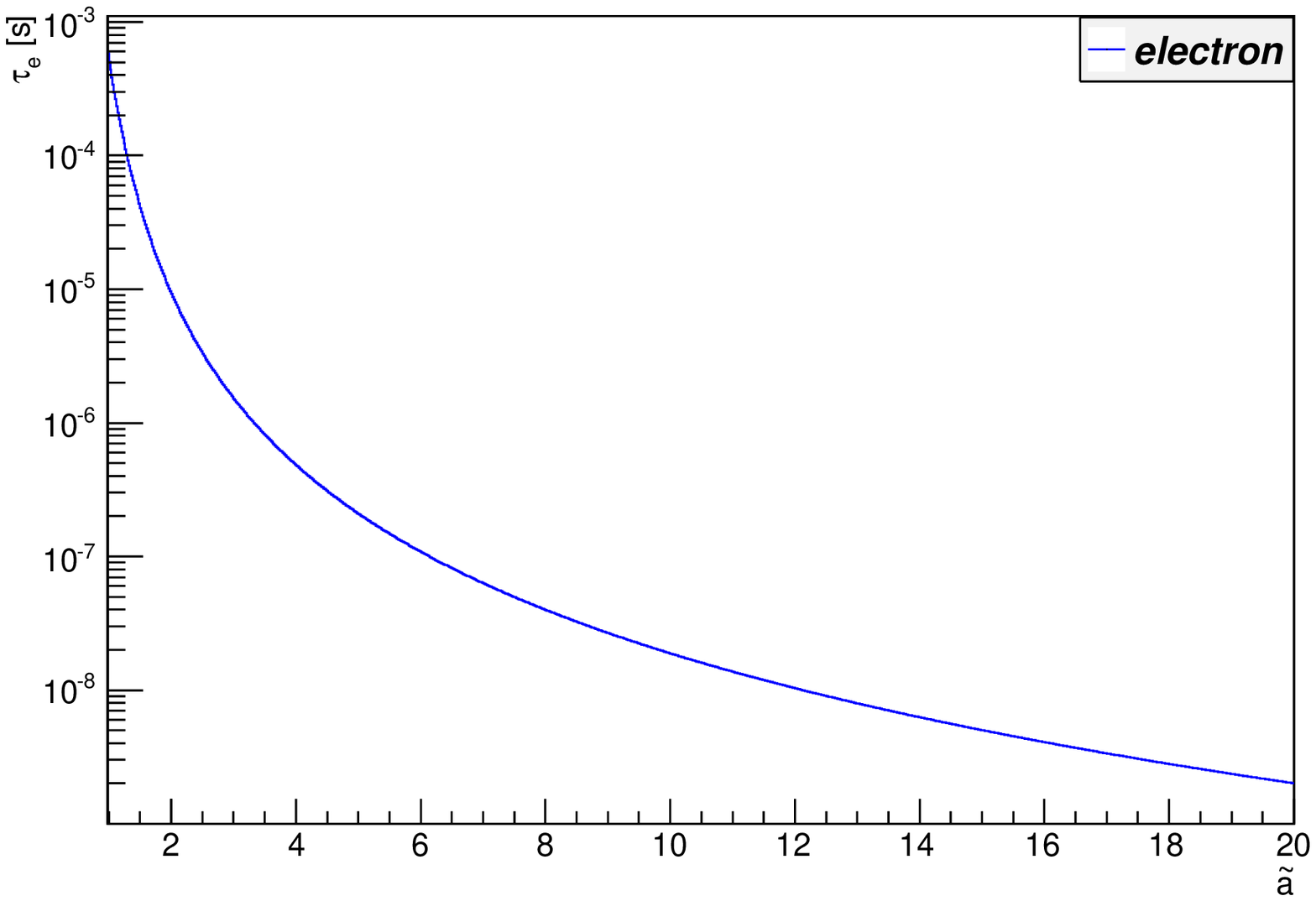}
\caption{The electron lifetime $\tau_{e} = 1/\Gamma_{e}$ as a function of $\tilde{a} = \frac{a}{m_{\m}}$.}
\end{figure}

This is a first estimate of the electron lifetime under the presence of uniform acceleration. A more accurate calculation would necessitate the inclusion of fermion fields as well as a weak or Fermi interaction Lagrangian. The use of fermions in the mathematically similar process of acceleration induced proton decay [6] yields a higher order polynomial of multiplicity in the decay rate due to additional factors of frequency in the fermionic normalization. This will not affect our result in the limit of low acceleration $a< m_{\m}$. For higher accelerations, the difference between the scalar and fermionic description would be a higher order polynomial of multiplicity.

This analysis can be further utilized to investigate the various decay chains of accelerated muons. In this investigation we will assume all final state products to be massless and are emitted into Minkowski space, i.e. $\Delta E_{R} = -m_{\m}$. This will allow us to get a better understanding of the overall conceptual properties of how the branching fractions of unstable particles change as a function of acceleration. Excluding any exotic or lepton number violating modes [13], there are three known decay channels for muons. These decay chains and their associated branching fractions are listed below:

\bqa
 &\;& \Gamma_{1}[\m \rightarrow e\bar{\nu}_{e}\nu_{\m}]: \;\;\;\;\;Br_{1} = 0.98599966 \non \\
 &\;& \Gamma_{2}[\m \rightarrow e \bar{\nu}_{e}\nu_{\m}\gamma]: \;\;\; Br_{2}= 0.014 \non \\
 &\;& \Gamma_{3}[\m \rightarrow e\bar{\nu}_{e}\nu_{\m}e\bar{e}]: \;\; Br_{3} = 0.000034.
\eqa

We have seen in the previous sections that the high acceleration limit favors the decay chain with the most final state products. Below we include the decay rate and lifetime plots of each decay channel, appropriately normalized to the inertial muon limit, for $n_{M}=3,4,5$ final states from Eq. (60) weighted by their associated branching fractions (see Figs. 11 and 12). The crossover from the primary channel to the secondary and then tertiary takes place at approximately $a\sim 4m_{\m} \sim 400 \; \text{MeV}$. We also include, for completeness, the various branching fractions as a function of proper acceleration given by

\bqe
Br_{i}(a) = \frac{Br_{i}\Gamma_{i}(a)}{\sum_{j} Br_{j}\Gamma_{j}(a)}.
\eqe

Rather than looking for direct evidence of acceleration induced decays it may be more experimentally tenable to measure these processes through the branching fractions of the decay chains and their dependence on proper acceleration (see Fig. 13). This may provide an easier method of discovering this or related phenomena.

\begin{figure}[H]
\centering  
\includegraphics[,scale=.6]{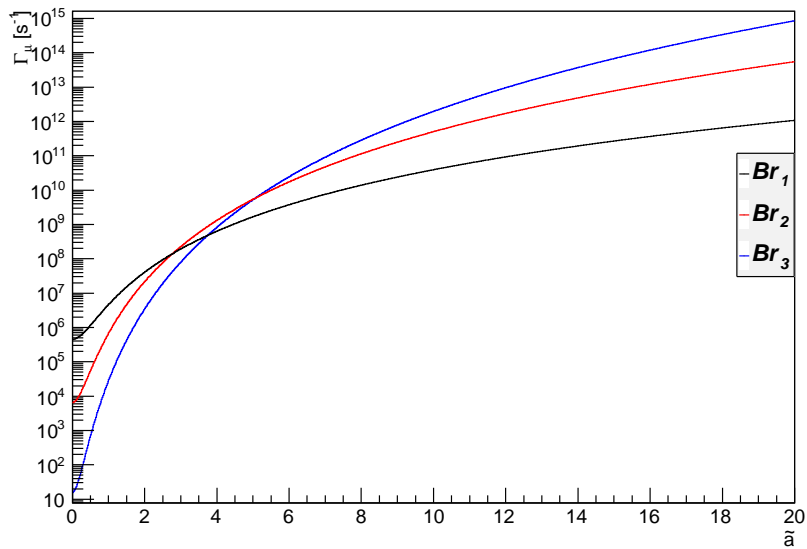}
\caption{The muon decay rates for the three known branching ratios, Eq. (69), as a function of $\tilde{a} = \frac{a}{m_{\m}}$.}
\end{figure}

\begin{figure}[H]
\centering  
\includegraphics[,scale=.6]{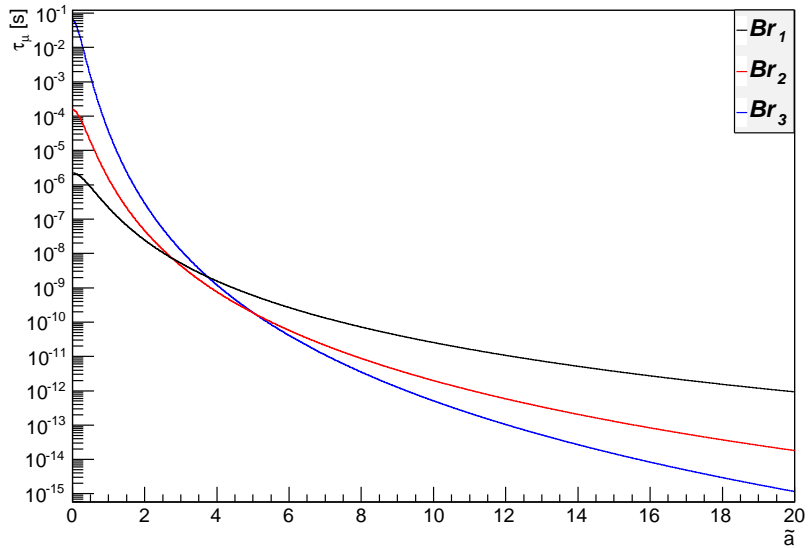}
\caption{The muon lifetimes $\tau_{\m} = 1/\Gamma_{\m}$ for the three known branching ratios, Eq. (69), as a function of $\tilde{a} = \frac{a}{m_{\m}}$.}
\end{figure}

\begin{figure}[H]
\centering  
\includegraphics[,scale=.6]{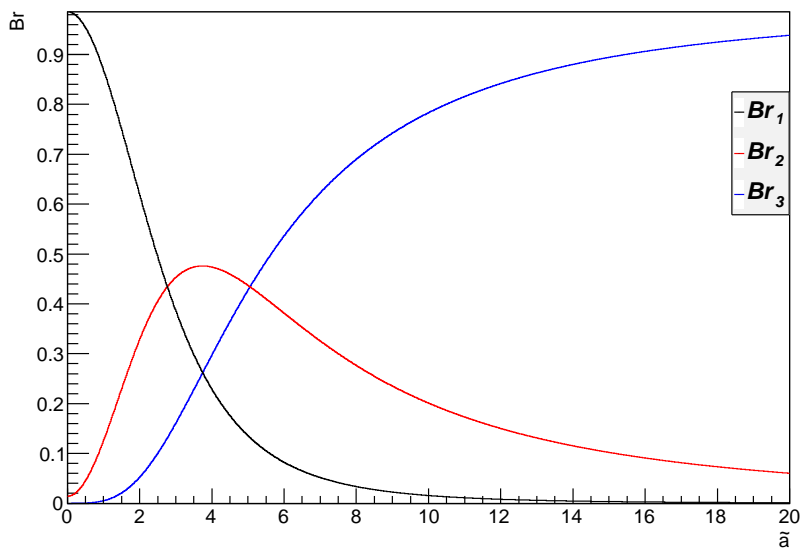}
\caption{The muon decay branching fractions, Eq. (70), as a function of $\tilde{a} = \frac{a}{m_{\m}}$.}
\end{figure}

\section{Conclusions}
In this paper we analyzed how acceleration affects the decay and excitation properties scalar fields with the simultaneous emission of an arbitrary number of final state products. We utilized methods of field operators and Unruh-DeWitt detectors to carry out the analysis. Generalized analytic results of $n$-particle multiplicities into both Rindler and Minkowski spacetimes were obtained. We included plots of all transition rates and lifetimes for various multiplicities as a function of acceleration and transition energy gaps. We found that high accelerations favor the decay chain with the most amount of Minkowski space final state products. The resultant formulas were applied to the muon-electron weakly interacting system and used to estimate the muon and electron lifetimes under acceleration. The evolution of the known branching fractions of muon decay under acceleration were also analyzed. Plots of all decay and excitation rates, proper lifetimes, and branching fractions were also included. 

\section*{Acknowledgments}
The author wishes to thank Luis Anchordoqui and Luiz da Silva for many valuable discussions as well as Ivan Agullo for proof reading this manuscript. This research was supported, in part, by the Leonard E. Parker Center for Gravitation, Cosmology, and Astrophysics and the University of Wisconsin at Milwaukee Department of Physics.

\goodbreak

\end{document}